\documentclass[
twocolumn,
tightenlines,
10pt,
longbibliography,
showpacs,
nofootinbib,
notitlepage,
superscriptaddress]{revtex4-2}

\usepackage{amsmath, amssymb, amsthm, braket, enumitem, dsfont, relsize, tabu}
\usepackage{makecell, graphicx, subfiles, multirow, wrapfig, hhline, microtype}
\usepackage[table, dvipsnames]{xcolor}
\usepackage[font=scriptsize]{subcaption}
\usepackage[font=scriptsize, justification=raggedright]{caption}
\usepackage[page]{appendix}
\usepackage[colorlinks, allcolors=black]{hyperref}
\usepackage[capitalize, nameinlink]{cleveref}
\usepackage{tikz}
\usepackage{pgfplots}
\pgfplotsset{compat=newest}

\definecolor{lightgray}{gray}{0.8}
\usetikzlibrary{quantikz}
\usetikzlibrary{shapes.geometric, arrows}
\usetikzlibrary{positioning}
\setlist[itemize]{leftmargin=10pt}

\DeclareCaptionLabelFormat{bold}{\textbf{(#2)}}
\makeatletter

    \def\CT@@do@color{%
      \global\let\CT@do@color\relax
            \@tempdima\wd\z@
            \advance\@tempdima\@tempdimb
            \advance\@tempdima\@tempdimc
    \advance\@tempdimb\tabcolsep
    \advance\@tempdimc\tabcolsep
    \advance\@tempdima2\tabcolsep
            \kern-\@tempdimb
            \leaders\vrule
                    \hskip\@tempdima\@plus  1fill
            \kern-\@tempdimc
            \hskip-\wd\z@ \@plus -1fill }
\makeatother

\begin{document}

\title{Coherent Ising Machines: The Good, The Bad, The Ugly}

\author{Farhad~Khosravi}
\affiliation{Irrevresible Inc., Sherbrooke, QC, Canada}

\author{Martin~Perreault}
\affiliation{Irrevresible Inc., Sherbrooke, QC, Canada}

\author{Artur~Scherer}
\affiliation{1QB Information Technologies (1QBit), Vancouver, BC, Canada}

\author{Pooya~Ronagh}
\thanks{{\vskip-10pt}{\hskip-8pt}Corresponding author: \href{mailto:pooya.ronagh@irreversible.tech}{pooya.ronagh@1qbit.com}\\}
\affiliation{Irrevresible Inc., Sherbrooke, QC, Canada}
\affiliation{1QB Information Technologies (1QBit), Vancouver, BC, Canada}
\affiliation{Institute for Quantum Computing, University of Waterloo, Waterloo, ON, Canada}
\affiliation{Perimeter Institute for Theoretical Physics, Waterloo, ON, Canada}

\date{\today}

\begin{abstract}
Analog computing using bosonic computational states is a leading approach to surpassing the computational speed and energy limitations of von Neumann architectures. But the challenges of manufacturing large-scale photonic integrated circuits (PIC) has led to hybrid solutions that integrate optical analog and electronic digital components. A notable example is the coherent Ising machine (CIM), that was primarily invented for solving quadratic binary optimization problems. In this paper, we focus on a mean-field interpretation of the dynamics of optical pulses in the CIM as solutions to Langevin dynamics, a stochastic differential equation (SDE) that plays a key role in non-convex optimization and generative AI. This interpretation establishes a computational framework for understanding the system’s operation, the computational role of each component, and its performance, strengths, and limitations. We then infer that the CIM is inherently a continuous state machine, capable of integrating a broad range of SDEs, in particular for solving a continuous global (or mildly constrained) optimization problems. Nevertheless, we observe that the iterative digital-to-analog and analog-to-digital conversions within the protocol create a bottleneck for the low power and high speed of optics to shine. This observation underscores the need for major advances in PIC technologies as we envision that fully analog opto-electronic realizations of such experiments can open doors for broader applications, and orders of magnitude improvements in speed and energy consumption.
\end{abstract}

\maketitle

\section{Introduction}

In recent years, the design of special-purpose bosonic computing architectures for optimization~\cite{mohseni2022ising,Stroev2023} and machine learning~\cite{shastri2021photonics,inagaki2025information} has gained increasing attention. This notably includes coherent optical networks~\cite{wang2013coherent, yamamoto2017coherent}, optical neural networks~\cite{hamerly2019large,xiao2021large}, and networks of light-matter particles~\cite{Kalinin2018}, which compete against physics-inspired heuristics such as thermal (and simulated) annealing~\cite{kirkpatric1983optimization, johnson1989optimization}, open system quantum dynamics such as quantum annealing~\cite{defalco1988numerical, apolloni1989quantum, kadowaki1998quantum, brooke1999quantum}, conventional local heuristic search algorithms~\cite{benlic2017breakout, prajapati2020tabu}, and high-performance spin-based Monte Carlo simulations~\cite{katzgraber2015seeking, aramon2019physics, takemoto20192}. Among these technologies, the coherent Ising machine has achieved impressively large experimental scales, solving Ising problems with hundreds of thousands of variables \cite{hamerly2019experimental, leleu2019destabilization, leleu2021scaling, reifenstein2021coherent, sankar2021benchmark, honjo2021spin}, however, the existence of a genuine computation time or energy consumption advantage has remained shrouded in doubt.

The well-known time-multiplexed coherent Ising machine (CIM) consists of a network of degenerate optical parametric oscillators (degenerate OPO, or DOPO) that are injected into a ring cavity and gradually pumped at a rate well above the bifurcation threshold~\cite{wang2013coherent, yamamoto2017coherent}. Such coherent optical networks can be realized either fully optically using optical delay lines (DL-CIM)~\cite{takata2015quantum, maruo2016truncated}, or with the assistance of a digital processing device, e.g., a field-programmable gate array (FPGA), in an iterative measurement-feedback procedure (MF-CIM) ~\cite{leleu2019destabilization, kako2020coherent}. While the fully optical DL-CIM can take advantage of fast optical clock speeds, it is difficult to scale up for arbitrarily structured optimization problems with all-to-all connectivity. In addition, the types of optimization problems solved using DL-CIM are limited by the optical interactions available. Since higher-order interactions are difficult to optically implement, solving higher-order optimization problems has appeared challenging. Indeed, solving optimization problems using DL-CIM has been limited to small problems with quadratic objective functions.

The MF-CIM, on the other hand, alleviates the challenges with connectivity and higher-order optimization problems by replacing the required complicated optical circuits with a digital processor which performs all the arithmetic subroutines. The MF-CIM has been used to solve optimization problems with thousands of variables using a single FPGA~\cite{mcmahon2016fully}. Although greatly beneficial in reducing the complexity and improving the scalability of the device, the digital processor acts as a bottleneck, preventing the full advantage of faster clock speed and lower energy consumption of optics to shine. This is one of the main themes of investigation in this paper. Moreover, the mainstream focus on binary, and even more restrictively quadratic, optimization problems means that realistic optimization problems (e.g., those comprising continuous variables) require costly recasting and discretizations. This results in very large Ising reformulations with ill-behaved energy landscapes that are difficult to optimize. These challenges have hindered the practical applicability of the CIM technology despite over a decade of commercial efforts. A broader (and more classical) view to the computational mechanisms of the operation of CIM is the second theme in our exposition. We do not consider the apparent absence of quantum computational means in our picture as a drawback of the CIM. On the contrary, we leverage this semi-classical interpretation both to pinpoint the important role of non-linear optics in this scheme, and to envision alternative high-value applications for it.

When solving binary optimization problems, CIMs begin an optimization process that produces classical computational states from the quantum states of light~\cite{yamamoto2020coherent}. In a low signal-loss setting, a product of coherent cat states of the form $|\alpha \rangle + |-\alpha \rangle$ is generated at slightly above the bifurcation threshold~\cite{kiesewetter2022coherent}. Iterative weak measurements then result in a gradual collapse of the coherent product state into a product of classical bits representing a low-energy state of the optimization problem. This quantum-to-classical transition has been named an ``exponential amplification'' of the low-energy states and superior to the ``linear amplitude amplification'' of Grover's search~\cite{yamamoto2020coherent}. However, one cannot consider the mere collapse of the coherent product state as quantum amplification. In what follows, we instead focus on the Langevin equations (the diffusion SDE) of the CIM \cite{lax1966quantum,lax1969quantum}, as the mechanism responsible for solving optimization problems. The CIM evolves its state according to a diffusion process generated by the iterative noise injection of its nonlinear crystal and the measurement process. Langevin dynamics is known to mix into low-energy states of the dynamical system's potential when the dynamical system possesses detailed balance~\cite{chiang1987diffusion,xu2017global,kurchan1998fluctuation,dal2021fluctuation,ritort2022fluctuation}. However, in the Langevin SDE of the CIM, the drift term deviates from the gradient of the objective function. This discrepancy, along with the non-convex landscape of the optimization problem, prevent the CIM from always succeeding in finding the global optimum even within the Ising solutions. Moreover, the diffusion coefficients can be manipulated in a way that the dynamical system under simulation violates physical dissipation-fluctuation relations.

In order to assess whether CIM's deviations from the simpler overdamped Langevin dynamics (OLD) plays any role in its performance as an optimizer, we demonstrate the on par asymptotic effectiveness of both dynamics in solving binary-variable (BV) optimization problems. Viewing the CIM as an energy-efficient and fast approximate integrator of Langevin dynamics opens doors for its usage in solving continuous non-convex optimization problems and generative AI inference (``the good''). Indeed, being intrinsically continuous, the analog amplitudes of the DOPO pulses can be used to represent the continuous variables of a continuous-variable (CV) optimization problem below the saturation threshold. The continuous readout of CIM solutions\footnote{While the coherent optical network used to solve the CV optimization problem is not solving an Ising problem, we still call such device a CIM.} was first explored in \cite{khosravi2022non}. In this paper, we further study the similarities and differences between the CIM and OLD dynamics, and use our observations to refute a critical need for preparing coherent cat states of light, casting doubts on `quantumness' of the computation performed by CIMs (``the bad''). On the other hand, by analyzing the energy consumption of the CIMs compared against digitally implemented devices, we show that hybrid optical-digital devices cannot perform better than digitally implemented OLD due to the bottleneck of digital data processing. We, therefore, conclude that the true advantage of optical acceleration of the computation will only be unlocked when the digital processor is removed from the architecture (``the ugly'').

This paper is organized as follows. In \cref{sec:langevin_and_con}, we show the SDEs governing the common CIMs and compare them with the OLD and its modified variants. In \cref{sec:the_good}, we compare the performance of the CIMs against OLD, and other heuristic binary solvers, when solving a simple non-convex optimization problem. In \cref{sec:the_bad}, we look at the time-to-solution (TTS) metric for three types of hardware (fully digital OLD solver, hybrid optical-digital MF-CIM, and fully optical DL-CIM) and compare their performance against each other. In \cref{sec:the_ugly}, we look at TTS and energy-to-solution (ETS) for these three types of hardware and investigate the merits of using analog optical devices for solving SDEs. We conclude with a brief discussion and an outlook on future research in \cref{sec:conclusion}.

\section{Semi-classical CIM Dynamics}
\label{sec:langevin_and_con}

In the continuous-time model of the \mbox{DL-CIM~\cite{takata2015quantum, maruo2016truncated}}, a system of SDEs describe the full quantum dynamics of the amplitudes of the DOPO pulses inside a cavity. This is an accurate quantum model of the CIM at least in the limit of high-finesse optical cavities, i.e., with low optical loss compared to the lifetime of the photons in the system~\cite{ng2022efficient}. In the positive \mbox{P-representation}, the random variables $c_i$ and $s_i$ representing the in-phase and quadrature-phase components of the signal field~\cite{Marandi2014} evolve according to the following SDEs:
\begin{equation}
\begin{split}
dc_i = & \left[\left(-1+p - c_i^2 - s_i^2\right)c_i - \lambda \partial_i f(c) \right]dt
\\ & + \frac{\zeta(t)}{A_s}\sqrt{c_i^2 + s_i^2 + \frac{1}{2}}\; dW_{i1} , \\
ds_i = & \left[\left(-1-p - c_i^2 - s_i^2\right)s_i - \lambda \partial_i f(s) \right]dt
\\ & + \frac{1}{\zeta(t)A_s}\sqrt{c_i^2 + s_i^2 + \frac{1}{2}}\; dW_{i2} .\\
\end{split}
\label{eq:sde_DL-CIM}
\end{equation}
Here the Wiener increments $dW_{i1}$ and $dW_{i2}$ are  independently sampled from an identical distribution, \mbox{$d W_{im}\sim \sqrt{dt}\, \mathcal{N}(0, 1)$,} hence injecting Gaussian noise of mean zero and variance of $dt$ into the dynamics. The value $A_s =(\gamma_p \gamma_s/2\kappa^2)^{1/2}$ is determined via the signal and pump decay rates, $\gamma_s$ and $\gamma_p$, respectively. The parameter $\kappa$ represents the parametric gain due to the second-order susceptibility of the nonlinear crystal. The parameter $\zeta(t)$ controls the variance of the injected noise during the computation, and is implemented by continually injecting a squeezed vacuum state into the open port of the beamsplitter that out-couples the optical pulses from the cavity into the delay-line network~\cite{mcmahon2016fully}. The first drift term represents three physical processes: the constant $-1$ represents a photon loss rate that is normalized to that of the optical cavity; the parameter $p$ represents the strength of the external pump field (e.g., a laser), normalized to the photon decay rate at the signal frequency $\gamma_s$, inducing an amplification of field strengths in the cavity during the process; and the term \mbox{$c_i^2 + s_i^2$} results from  the nonlinear self-interaction induced by a nonlinear crystal (e.g., a periodically poled lithium niobate crystal (PPLN))~\cite{wang2013coherent}.

The goal is to optimize the differentiable real-valued function $f: \mathbb R^n \to \mathbb R$. We use $\partial_i$ to denote the $i$-th partial derivative $\partial_i= \partial/\partial c_i$. For example, if $f(c)= \sum_{i, j=1, \ldots, N} \xi_{ij} c_i c_j$ is a quadratic function, then the $i$-th drift term is proportional to \mbox{$\partial_i f(c)= 2\sum_{j=1}^N \xi_{ij} c_j$,} which may be realized physically via two-mode interactions. We note that the drift term in~\cref{eq:sde_DL-CIM} may not be experimentally realizable for arbitrary differentiable functions $f$ as it requires implementing higher-order optical interactions, which is a challenge being tackled in current research~\cite{yanagimoto2020engineering}. Therefore, at least for the fully optical implementation of the CIM, we shall restrict our attention to the case of quadratic functions, as two-mode interactions of programmable strength can be realized using optical delay lines and beam splitters. With this in mind, we refer to the system of SDEs~\eqref{eq:sde_DL-CIM} as {\textit{DL-CIM dynamics}}.

In the MF-CIM, assuming the DOPO pulses maintain a Gaussian distribution throughout the evolution~\cite{leleu2019destabilization, kako2020coherent}, the associated mean values ($\mu_i$) and variances ($\sigma_i$) follow the slightly different set of SDEs~\cite{shoji2017quantum}
\begin{equation}
\begin{split}
d\mu_i = & \left[-(1+j) + p - g^2 \mu_i^2\right] \mu_i\, dt\\
& - \lambda \partial_i f(\tilde\mu)\, dt
+ \sqrt{j}(\sigma_i - 1/2) dW_i ,\\
d\sigma_i = & 2\left[ -(1+j) + p - 3g^2 \mu_i ^2  \right]\sigma_i\, dt  \\
& - 2j(\sigma_i - 1/2)^2\, dt + \left[(1+j) + 2g^2 \mu_i ^2 \right] dt.
\end{split}
\label{eq:sde_MF-CIM}
\end{equation}
Here, $j$ is the normalized continuous measurement strength, and $g$ is the normalized second-order nonlinearity coefficient. We refer to this system of SDEs as {\textit{MF-CIM dynamics}}. Note that, unlike in the previous systems of SDEs, the gradient field of $f$ defined by~\cref{eq:sde_MF-CIM} is evaluated at a measured mean-field amplitude vector $\tilde\mu$ which differs from the instantaneous mean-field amplitude within the cavity by a random vector representative of the uncertainty of the continuous quantum measurements:
\begin{equation}
\tilde\mu= \mu + \sqrt{\frac{1}{4j}}\, \frac{dW}{dt}.
\end{equation}
The MF-CIM scheme can be realized experimentally for arbitrary differentiable functions $f$, since the feedback term $-\lambda \partial_i f (\tilde \mu)$ is calculated via a digital processor, for example, an FPGA, GPU, an application-specific integrated circuit (ASIC), or perhaps via on-chip photonics.

The DL-CIM and MF-CIM dynamics described above are both Langevin equations of dissipative physical systems of the general form
\begin{equation}
\label{eq:sde_langevin}
dx= b(x, t)\, dt -\nabla f(x)\, dt + \sigma(x, t)\, dW,
\end{equation}
where $x$ represents either of the quantum or mean-field degrees of freedom $(c,s)$ or $(\mu, \sigma)$. Here $b(x, t)$ is a component-wise external force driving the dynamics of each DOPO controlled by the DOPO pump rate. The variable $W$ is a standard Wiener process realized by weak measurements of the optical modes either via photon loss in a highly dissipative cavity, particularly in a DL-CIM, or via out-coupling and homodyne detection in a MF-CIM. Finally, $\sigma (x, t)$ represents the diffusion rate, and is determined by the cavity finesse and optical nonlinearity of the DL-CIM, or by controlling the out-coupling rate of beam splitters in the MF-CIM.

\begin{figure}[!t]
\includegraphics[width=1.0\linewidth]{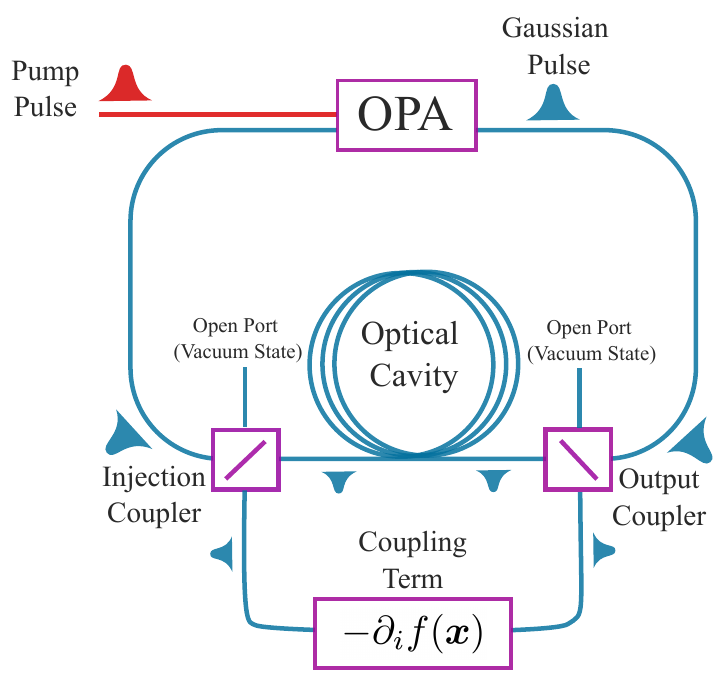}
\caption{Schematic of the architecture of common CIMs. Squeezed coherent states of light are generated and amplified using an OPA element that inclues a nonlinear crystal pumped by a laser that, together with a ring cavity, constitutes a DOPO. Information is encoded in time-multiplexed oscillations of the resonator, which are coherently amplified each time they pass through the OPA element. The coupling term is implemented by taking a portion of each pulse using an output coupler, calculating the gradient of the objective function using either a delay-line or measurement-feedback scheme, and feeding the result into each optical pulse using an injection coupler. These couplers, in general, can be variable beam splitters.}
\label{fig:ccvm-schematic}
\end{figure}

To explore the dynamics of CIMs as heuristics for the global optimization of $f$, we compare their performance against simpler Langevin equations with well-understood convergence properties. The first equation is the \emph{overdamped} Langevin dynamics (OLD),
\begin{equation}
dc = - \lambda \nabla f(c)\, dt + \sigma d W,
\label{eq:overdamped_langevin}
\end{equation}
the convergence rate of which to the Gibbs state $\propto \exp(-\beta f)$ with $\beta = 2\lambda/\sigma^2$ has been thoroughly studied in mathematics and computer science~\cite{roberts1996exponential, xu2017global, raginsky2017non}. To incorporate the effect of the pump field, we may consider the following modification of OLD,
\begin{equation}
dc_i = (-1 + p - c_i^2)c_i\, dt - \lambda \partial_i f(c)\, dt + \sigma d W_i,
\label{eq:pumped_langevin}
\end{equation}
which mixes into the Gibbs state of a perturbed potential
\begin{equation}
c \mapsto f(c) + \frac{1-p}2 \sum_{i} c_i^2 + \frac{1}4 \sum_i c_i^4.
\label{eq:perturbed-potential}
\end{equation}
Reference \cite{khosravi2022non} refers to \cref{eq:pumped_langevin} as the \emph{pumped} Langevin dynamics. Note that in both \cref{eq:overdamped_langevin,eq:pumped_langevin} the drift and diffusion coefficients $\sigma$ and $\lambda$ may vary through the evolution time toward their long-time final values $\sigma(T)$ and $\lambda(T)$ for large $T\gg 0$.

\cref{fig:ccvm-schematic} shows the general experimental scheme used to implement the DL- and MF-CIMs~\cite{yamamoto2017coherent}. Unconventionally, the device may operate below the saturation threshold. The key element of the architecture is a DOPO consisting of an optical resonator in the form of a ring cavity along with a nonlinear optical crystal pumped by a laser. The computational state of the machine is stored in the time-multiplexed oscillations of this resonator. Squeezed states are generated and continually amplified using optical parametric amplification (OPA) based on spontaneous parametric down-conversion. The resulting time-multiplexed pulses are coherently and iteratively coupled with the gradient of the objective function, providing a descent direction for the dynamics. The gradient can be implemented by either a delay-line or measurement-feedback scheme. The diffusion process is realized by the injection of quantum noise during the OPA process, as well as the quantum measurement process in the case of the MF-CIM.

\subsection{Ablation of the nonlinear crystal}

If the nonlinear crystal is removed from the experiment of \cref{fig:ccvm-schematic}, the in-phase and quadrature-phase components $c_i$ and $s_i$ in \cref{eq:sde_DL-CIM} are decoupled and the dynamics reduces to
\begin{equation}
\begin{split}
dx_i = & -\lambda \partial_i \left[\frac{x_i^2}{2 \lambda} + f(x) \right]dt
\end{split}
\label{eq:sde_DL-CIM_ablated}
\end{equation}
in the case of the DL-CIM, a simple and deterministic gradient descent on a regularized landscape deviating from the desired potential $f$ itself. However, interestingly, in the case of MF-CIM, the removal of the OPA will result in
\begin{equation}
\begin{split}
d\mu_i =& -(1+j) \mu_i\, dt - \lambda \partial_i f(\tilde\mu)\, dt \\
&+ \sqrt{j}(\sigma_i - 1/2) dW_i ,\\
d\sigma_i =& (\sigma_i - 1/2) (-2 - j - 2j\sigma_i) dt.
\end{split}
\label{eq:sde_MF-CIM_ablated}
\end{equation}
Since $\sigma_i = 1/2$ and $\sigma_i= -1/j - 1/2 < 0$ are the only steady state solutions of $\sigma_i$ and the negative solution is not physical, the dynamics converges to $\sigma_i = 1/2$ and the surviving $\mu$ dynamics converges asymptotically to
\begin{equation}
\begin{split}
d\mu_i &= -(1+j) \mu_i\, dt - \lambda \partial_i f(\tilde\mu)\, dt
\end{split}
\label{eq:sde_MF-CIM_ablated_mu_tilde}
\end{equation}
or equivalently,
\begin{equation}
\begin{split}
d \mu_i =& -(1+j) \mu_i dt - \lambda \partial_i f( \mu)\, dt \\
&+ (1+j) \sqrt\frac{1}{4j} dW_i + \sqrt\frac{1}{4j} \frac{d^2W_i}{dt^2} dt.
\end{split}
\label{eq:sde_linear-MF-CIM}
\end{equation}
This stochastic process includes not only the Wiener increment but also the time-derivative of white noise which cannot be interpreted in the usual It\^o sense but it can be rigorously studied using the theory of \emph{generalized} stochastic processes~\cite{hida1980brownian}. In this framework, \cref{eq:sde_linear-MF-CIM} means that for every real-valued compactly supported test function $\varphi(t)$ we expect the following integral equation to be satisfied:
\begin{equation}
\begin{split}
\int_{-\infty}^\infty \mu_i (t)\, \dot \varphi(t) \, dt
&= \int_{-\infty}^\infty \varphi(t)
\Big(
(1 + j)\, \mu_i
+ \lambda\, \partial_i f(\tilde{\mu})
\Big) dt \\
&+ (1 + j)\, \sqrt{\frac{1}{4j}}\, \int_{-\infty}^\infty
W(t) \dot \varphi(t)\, dt \\
&- \sqrt{\frac{1}{4j}}\, \int_{-\infty}^\infty
W (t) \ddot \varphi(t)\, dt.
\end{split}
\end{equation}
Similar to white noise, its derivative is a (generalized) Gaussian process. However, the covariance operator of white noise is the identity:
\begin{equation}
\mathbb{E}\left[ \dot{W} (\varphi) \dot{W} (\psi) \right]
= \int_{\mathbb{R}} \varphi(t)\,\psi(t)\,dt
= \langle \varphi, \psi \rangle,
\end{equation}
consistent with the martingale properties of the Wiener process, while the covariance operator of the derivative is $-d^2/dt^2$:
\begin{equation}
\mathbb{E}\left[ \ddot{W} (\varphi) \ddot{W} (\psi) \right]
= -\int_{\mathbb{R}} \ddot{\varphi}(t)\,{\psi}(t)\,dt
= \left\langle -\frac{d^2}{dt^2}\varphi,\psi \right\rangle.
\end{equation}
This means that the noise of process \eqref{eq:sde_linear-MF-CIM} does not have locally independent increments as expected from white noise. Further details can be found in \cref{sec:app-distributions}.

We name an MF-CIM that is lacking the nonlinear crystal a \emph{linear} MF-CIM, and abbreviate it as LMF-CIM. In summary, the absence of the nonlinear crystal completely eliminates the source of noise in the fully optical DL-CIM, but, interestingly, in the case of MF-CIM, the ablated experiment is still stochastic despite the variance of the signal field converging to a steady value. Note that since the spurious second-order noise discussed here is caused by homodyne detections, it is present even in the MF-CIMs that include the nonlinear crystal (see \cref{eq:sde_MF-CIM}).

\section{Prospects for Broad Utility\\(The Good)}
\label{sec:the_good}

\begin{figure}[t]
  \includegraphics[width=0.75\linewidth]{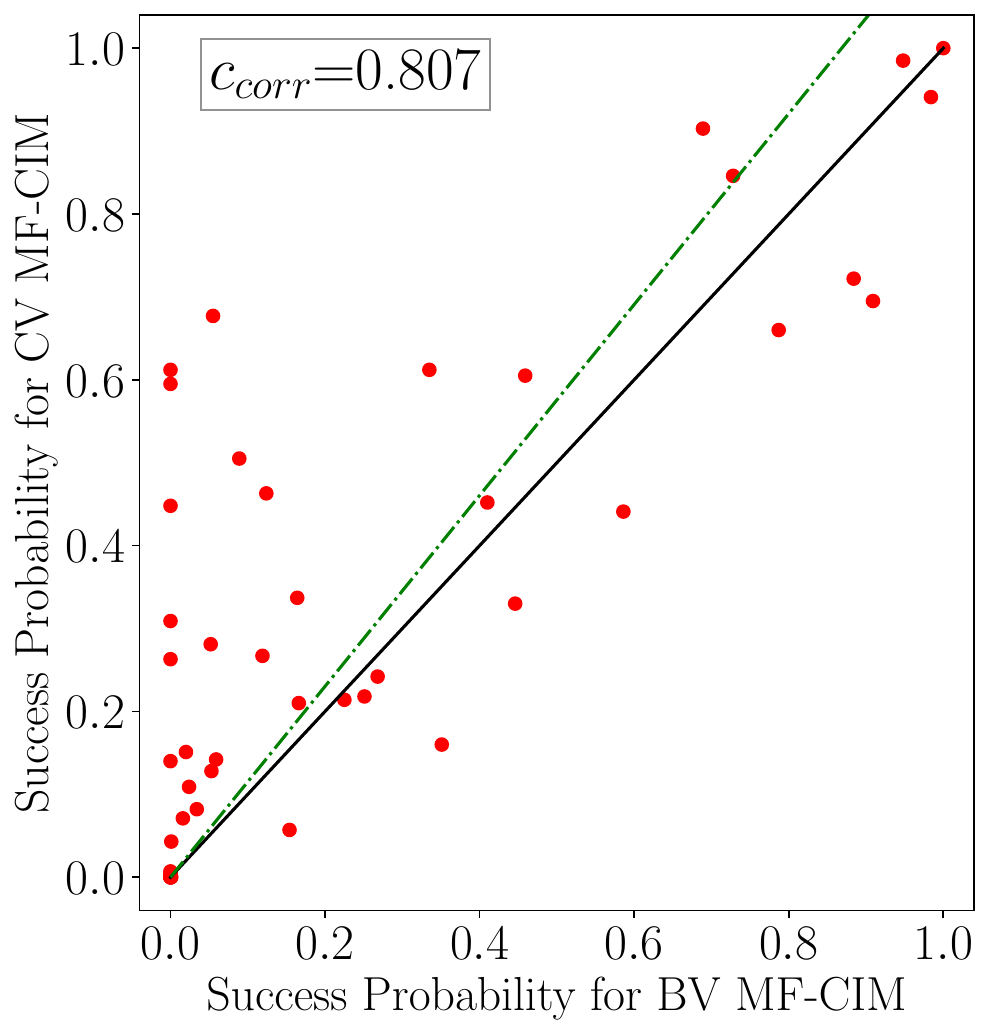}
  \caption{Correlation plot comparing MF-CIM's performance with CV and BV readouts when solving the CV BoxQP problem instances. This shows that a significant number of problem instances in the benchmark set require CV readouts for success even within the 0.1\% optimality gap. \label{fig:CV_vs_BV_MF-CIM}}
\end{figure}

\begin{figure*}[!ht]
  \subfloat[\label{fig:corr_MF-CCVM_vs_Langevin}]{\includegraphics[scale=0.31]{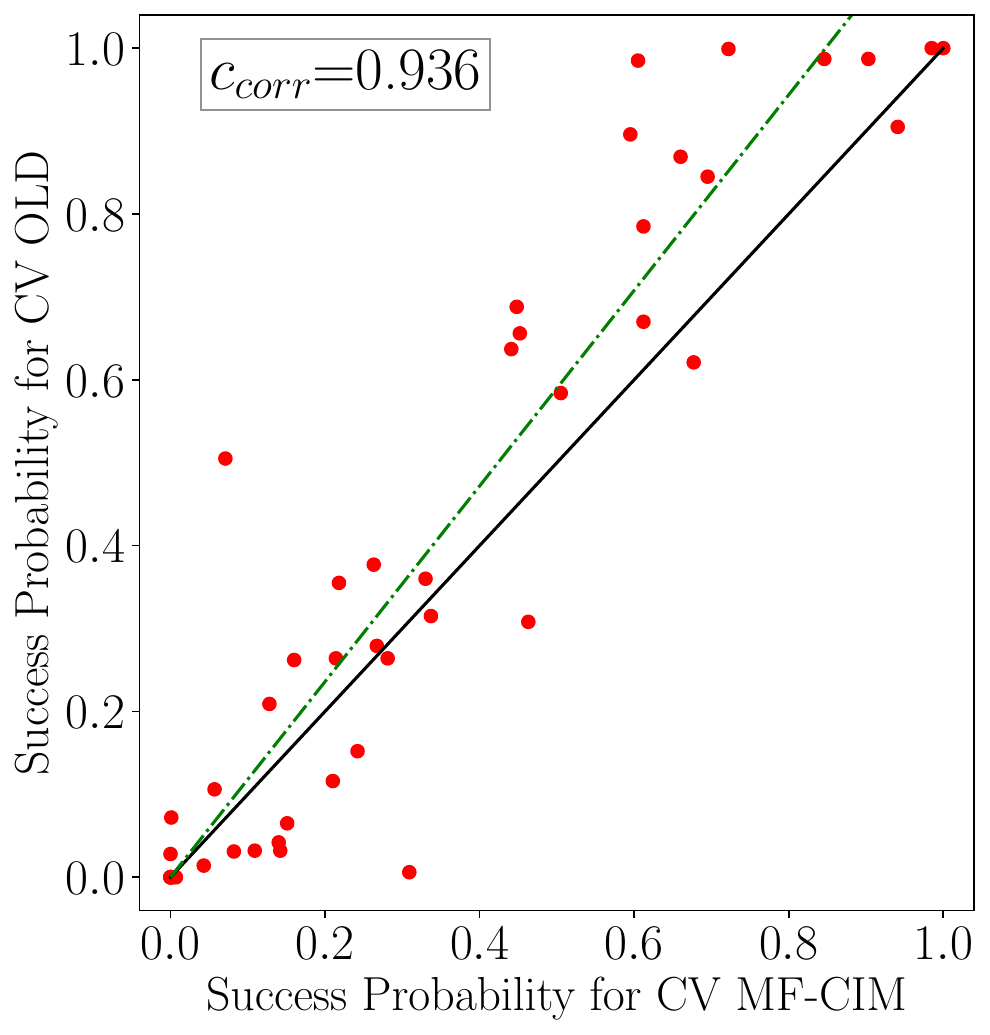}}\hspace{6mm}
  \subfloat[\label{fig:corr_MF-CCVM_vs_MF-CCVM_linear}]{\includegraphics[scale=0.31]{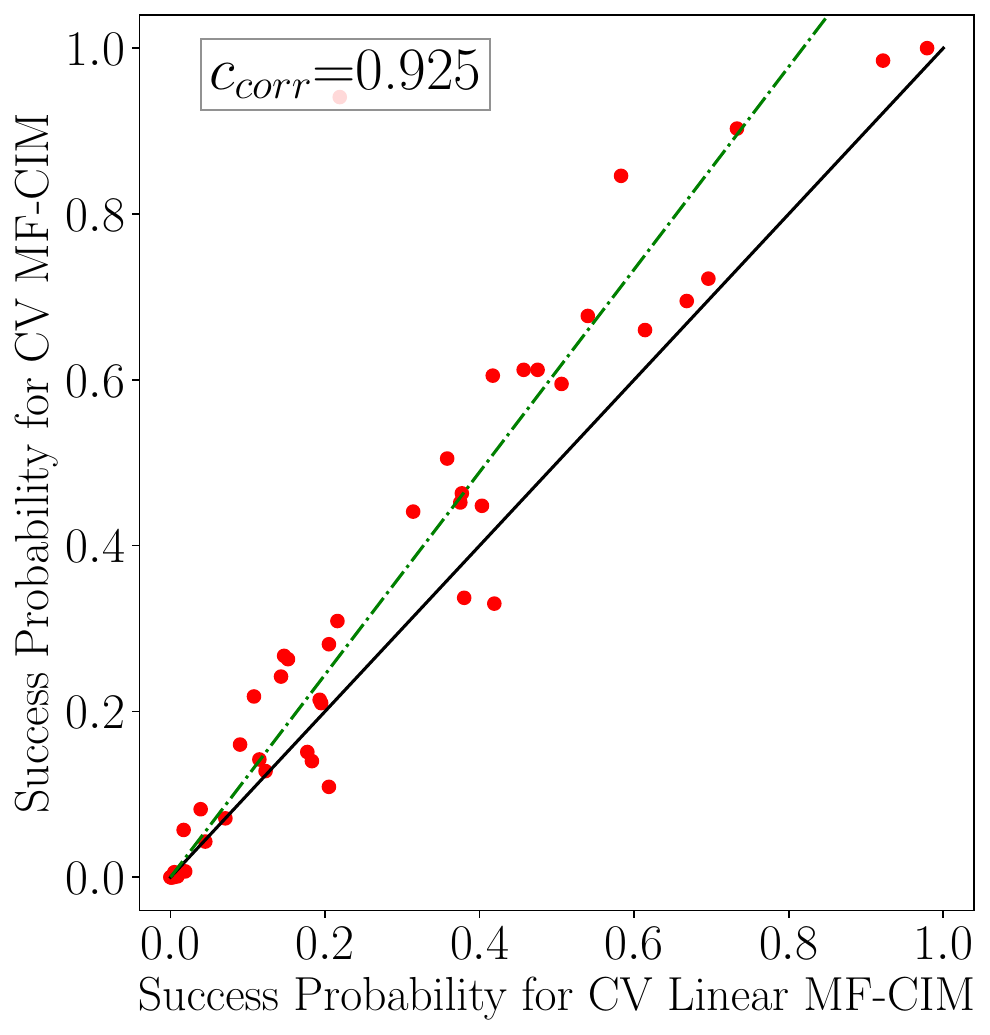}}\hspace{6mm}
  \subfloat[\label{fig:corr_MF-CCVM_linear_vs_gradient-descent}]{\includegraphics[scale=0.31]{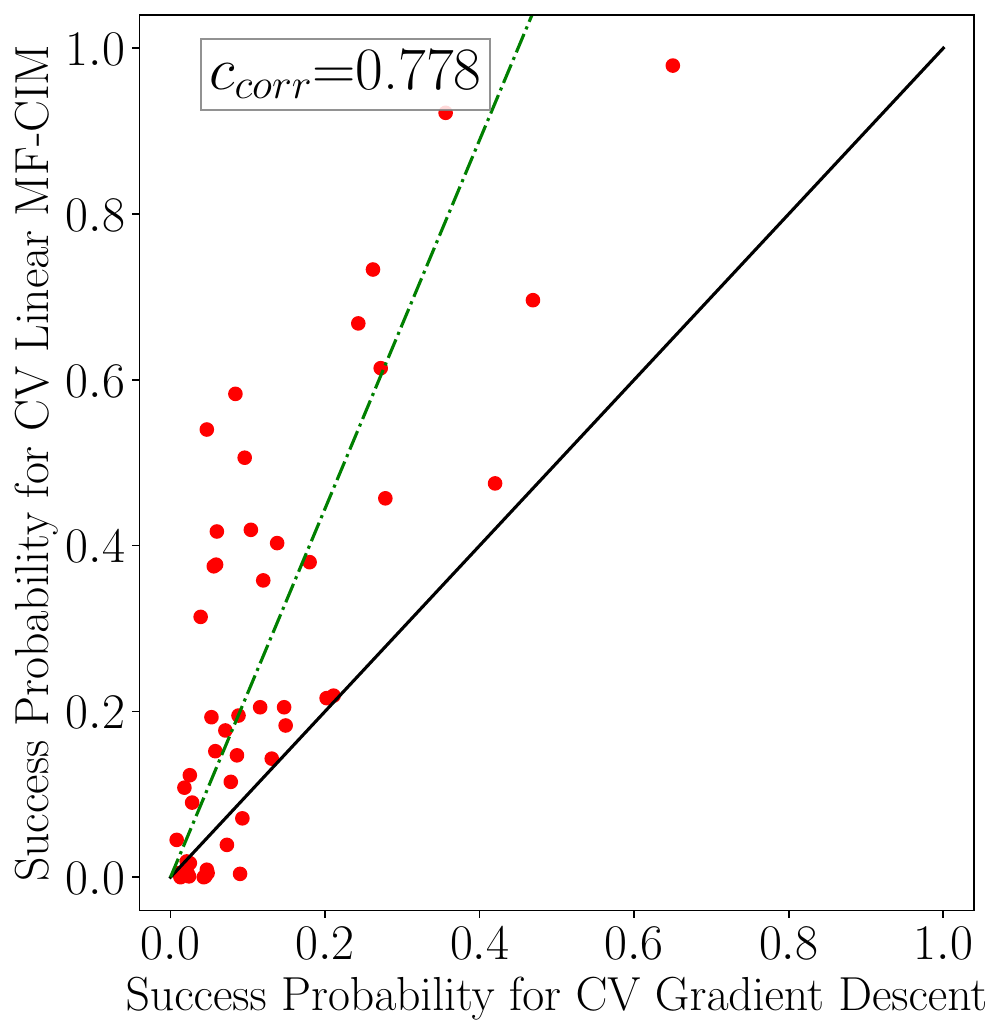}}\\
  \subfloat[\label{fig:corr_SA_vs_BLangevin}]{\includegraphics[scale=0.31]{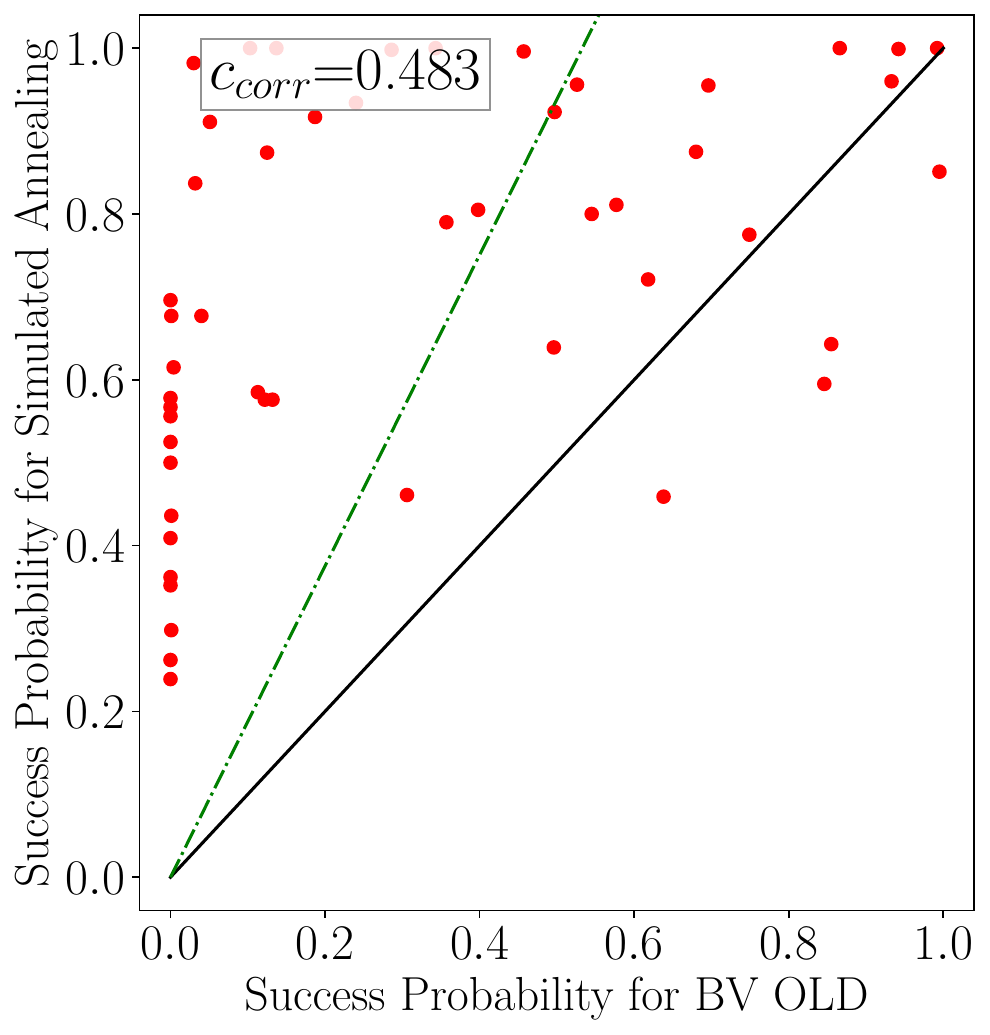}}\hspace{6mm}
  \subfloat[\label{fig:corr_BLangevin_vs_MF-CIM}]{\includegraphics[scale=0.31]{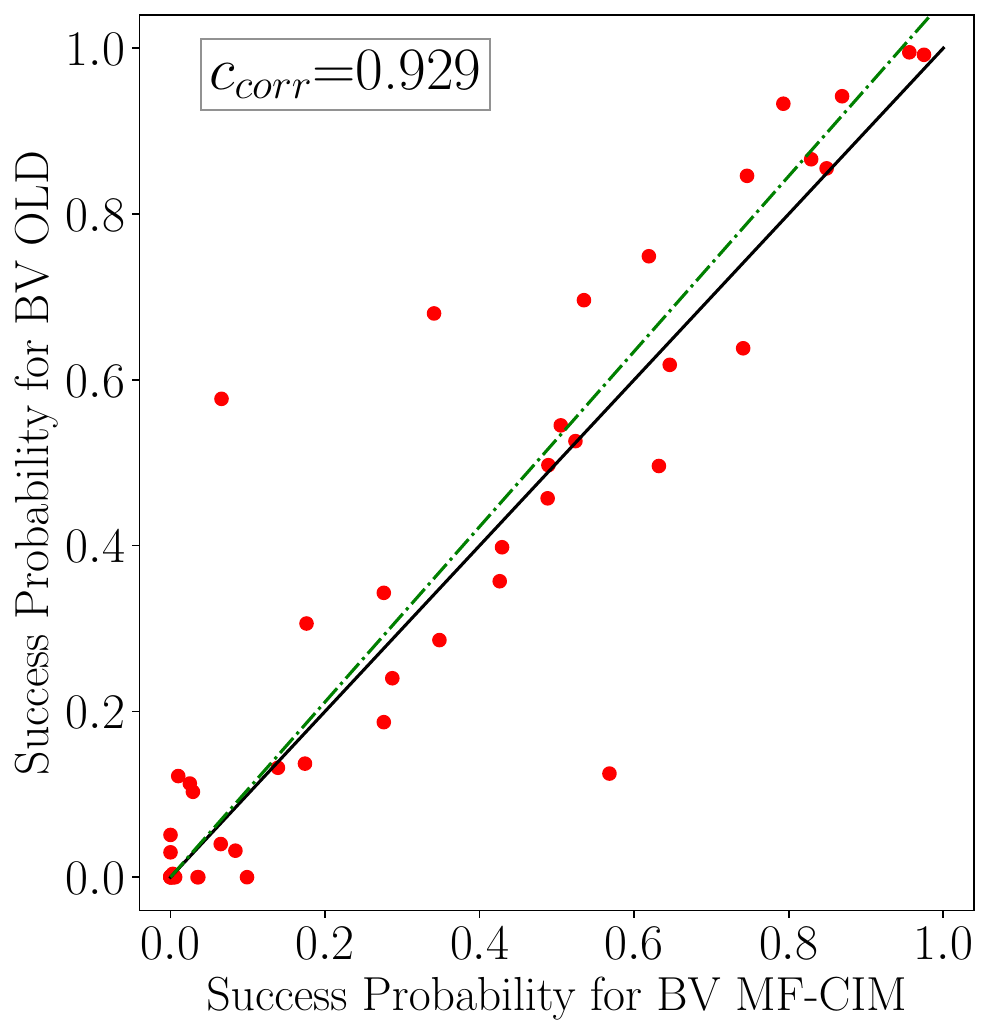}}\hspace{6mm}
  \subfloat[\label{fig:corr_MF-CIM_vs_SD}]{\includegraphics[scale=0.31]{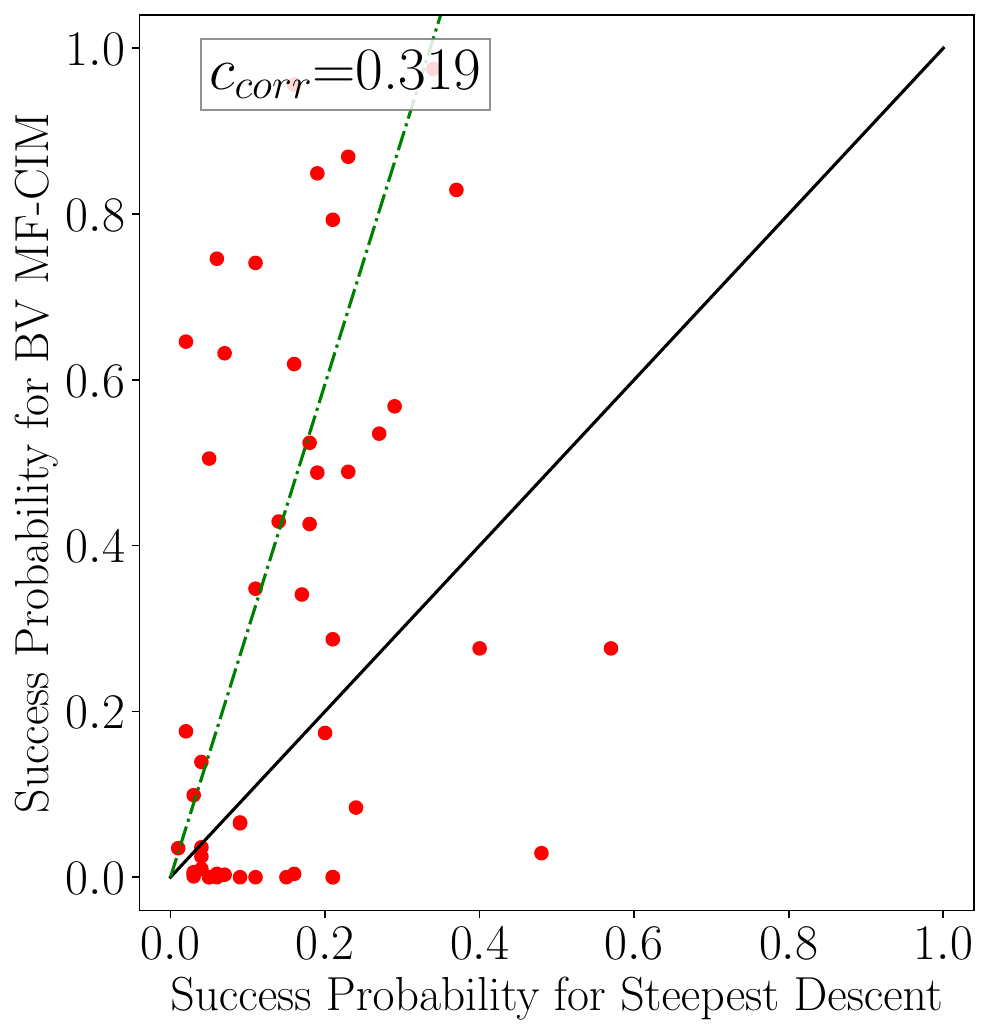}}
  \caption{Correlations between MF-CIM and classical heuristics when solving 50 BoxQP problem instances of size $70$. (a) CV-OLD against the CV implementation of MF-CIM, (b) CV-MF-CIM (with the nonlinear crystal) versus CV-LMF-CIM (no nonlinear crystal), and (c) CV-LMF-CIM versus CV gradient descent method with random initialization. Overall CV-OLD outperforms CV-MF-CIM, which outperforms CV-LMF-CIM, which in turn outperforms gradient descent. (d) Simulated annealing (SA) versus BV-OLD, (e) BV-OLD against BV-MF-CIM, and (f) BV-MF-CIM versus the steepest descent solver. The CV BoxQP problem was solved for (a), (b), and (c), whereas the BV problem with the same objective function was solved for (d), (e), and (f). The dashed green line is a linear regression of the data points intercepting the origin providing a quantitative aid for comparing pairs of solvers. The Pearson correlation coefficient of each pair of solvers is indicated as $c_\text{corr}$. An iteration number of $4000$ was used for all the OLD and MF-CIM methods and no post-processing steepest descents were performed. The success probabilities shown here indicate the results that are within $0.1\%$ gap of the optimum solutions found using a global optimizer (see \cref{sec:the_bad}).
  \label{fig:correlations}}
\end{figure*}

We now compare the performance of various CIMs against classical heuristics in solving a prototypical non-convex optimization problem, known as the {\em box-constrained quadratic programming} (BoxQP) problem~\cite{angelis1997quadratic, khosravi2022non}. This is the simplest non-convex optimization problem one can tackle, yet it is NP-hard~\cite{murty1985some} and no reasonably efficient classical algorithms for solving it exist~\cite{burer2009globally, bonami2018globally, vries2022tight}. Further information about the BoxQP problem, our construction of the benchmark instances, and how they are solved using the dynamics of CIMs are provided in \cref{sec:solvingbyCIM}.

We study the BV and CV cases of the same problem by simply converting the original continuous variables into binary ones. In what follows we use BV- and CV- prefixes to indicate a solver with continuous or binary readouts. The BV solvers are identical to their CV counterparts with the difference that the final results are rounded to binary values $0$ and $1$ in post-processing. \Cref{fig:CV_vs_BV_MF-CIM} compares the performances of CV-MF-CIM and BV-MF-CIM when solving the CV BoxQP problem instances. That is, if the solution is fractional the BV-MF-CIM fails. This test probes how many of our generated instances require fractional readout despite the tolerated optimality gap. The better performance of CV-MF-CIM indicates that MF-CIMs are capable of solving CV problems natively.

\Cref{fig:correlations} shows the correlations between MF-CIM variants, the OLD, and simple gradient and steepest descent algorithms. The correlation plots in \cref{fig:correlations} are ordered by decreasing solver performance from left to right, both for the CV (top row) and the BV (bottom row) solvers. The slope of the linear regression with the intercept fixed at the origin (green dashed) serves as a visual comparator for each pair of solvers. These results clearly indicate a large correlation (a Pearson correlation coefficients of greater than $\sim\! 0.8$) between various CIMs and OLD variants indicative of the similarities between the two dynamics. More specifically, \cref{fig:corr_MF-CCVM_vs_Langevin} shows the correlations in the success probabilities when solving the CV BoxQP problem using the CV-OLD versus the CV-MF-CIM solver. Next, \cref{fig:corr_MF-CCVM_vs_MF-CCVM_linear} compares the CV-MF-CIM solver against CV-LMF-CIM following the dynamics of \cref{eq:sde_linear-MF-CIM}, and \cref{fig:corr_MF-CCVM_linear_vs_gradient-descent} compares the performance of CV-LMF-CIM against a simple gradient descent with random initialization. Furthermore, for the task of binary optimization on the same problem instances, we have shown the correlation between simulated annealing (SA)~\cite{kirkpatric1983optimization}, the BV readout of OLD (BV-OLD), BV-MF-CIM, and the steepest descent (SD)~\cite{fliege2000steepest} algorithms in \cref{fig:corr_SA_vs_BLangevin,fig:corr_BLangevin_vs_MF-CIM,fig:corr_MF-CIM_vs_SD}, respectively.

Another interesting observation is that both CV-OLD and CV-MF-CIM only slightly outperform the CV-LMF-CIM solver, while CV-LMF-CIM is still by far better than (the deterministic) gradient descent. On the one hand, this shows that the shot noise of homodyne detection in the fully linear optical setup is already a good source for stochastic exploration of the optimization landscape. But nonetheless, the shot-noise driven dynamics of \cref{eq:sde_linear-MF-CIM} appears to be less favourable than the nonlinear optical setting of \cref{eq:sde_MF-CIM}. We leave it to future studies to investigate which factors among the time correlations in the derivative of white noise, the rigid dynamics of the signal variance, and the LMF-CIM’s inability to compensate for photon and measurement losses contribute most to this inferior performance.

Small correlation values of about $0.5$ between the BV-OLD or BV-MF-CIM solvers against both SA and steepest descent algorithms, as shown in \cref{fig:corr_SA_vs_BLangevin,fig:corr_MF-CIM_vs_SD}, also attest that dynamics of MF-CIM and OLD are fundamentally different from both the Metropolis-Hastings rejection sampling and the deterministic steepest descent algorithms (see also \cref{sec:effect_of_noise} for additional experiments). These results show that while ingenious quantum engineering plays an important role in the architecture of the CIMs, they ultimately realize classical computing schemes that approximately integrate the OLD and can solve generic BV and CV non-convex optimization problems with performance similar to that of OLD.

\begin{figure*}[t]
  \centering
  \hspace{-1.1cm}
  % -----------------------------
  % Subfigure 1: Convex surface
  % -----------------------------
  \begin{subfigure}[b]{0.28\textwidth}
    \centering
    \begin{tikzpicture}
      \begin{axis}[
        width=6cm,
        view={45}{45},
        domain=-1:1,
        y domain=-1:1,
        samples=35,
        samples y=35,
        xlabel=$x$,
        ylabel=$y$,
        zlabel=$z$,
        y dir=reverse,
        colormap/hot,  % << hot colormap for all
        title={{\bf (a)\ } $z = x^2 + y^2 - 0.25x - 0.25y$}
      ]
        \addplot3[surf,opacity=0.8]
          {x^2 + y^2 - 0.25*x - 0.25*y};
        \addplot3+[mark=*,color=red] coordinates {(0.125, 0.125, 0)};
      \end{axis}
    \end{tikzpicture}
  \end{subfigure}
  \hspace{0.7cm}
  % -----------------------------
  % Subfigure 2: Saddle surface
  % -----------------------------
  \begin{subfigure}[b]{0.28\textwidth}
    \centering
    \begin{tikzpicture}
      \begin{axis}[
        width=6cm,
        view={45}{45},
        domain=-1:1,
        y domain=-1:1,
        samples=35,
        samples y=35,
        xlabel=$x$,
        ylabel=$y$,
        zlabel=$z$,
        y dir=reverse,
        colormap/hot,
        title={{\bf (b)\ } $z = x^2 - y^2 - 0.25x - 0.25y$}
      ]
        \addplot3[surf,opacity=0.8]
          {x^2 - y^2 - 0.25*x - 0.25*y};
        \addplot3+[mark=*,color=red] coordinates {(0.125, 1, -1.265625)};
      \end{axis}
    \end{tikzpicture}
  \end{subfigure}
  \hspace{1cm}
  % -----------------------------
  % Subfigure 3: Concave surface
  % -----------------------------
  \begin{subfigure}[b]{0.28\textwidth}
    \centering
    \begin{tikzpicture}
      \begin{axis}[
        width=6cm,
        view={45}{45},
        domain=-1:1,
        y domain=-1:1,
        samples=35,
        samples y=35,
        xlabel=$x$,
        ylabel=$y$,
        zlabel=$z$,
        y dir=reverse,
        colormap/hot,
        title={{\bf (c)\ } $z = -x^2 - y^2 - 0.25x - 0.25y$}
      ]
        \addplot3[surf,opacity=0.8]
          {-(x^2) - y^2 - 0.25*x - 0.25*y};
        \addplot3+[mark=*,color=red] coordinates {(1, 1, -2.5)};
      \end{axis}
    \end{tikzpicture}
  \end{subfigure}

  \caption{Example BoxQP optimization landscapes in two variables $x$ and $y$. The global minimum of each optimization problem is shown via a red dot. (a) A convex surface with both $x$ and $y$ optimal values being fractional, $r= 1.0$. (b) A saddle surface, showing an extreme optimal value and a fractional one, $r= 0.5$. And (c) a concave surface with extreme optimal values in both variables, $r= 0.0$.}
  \label{fig:surfaces}
\end{figure*}
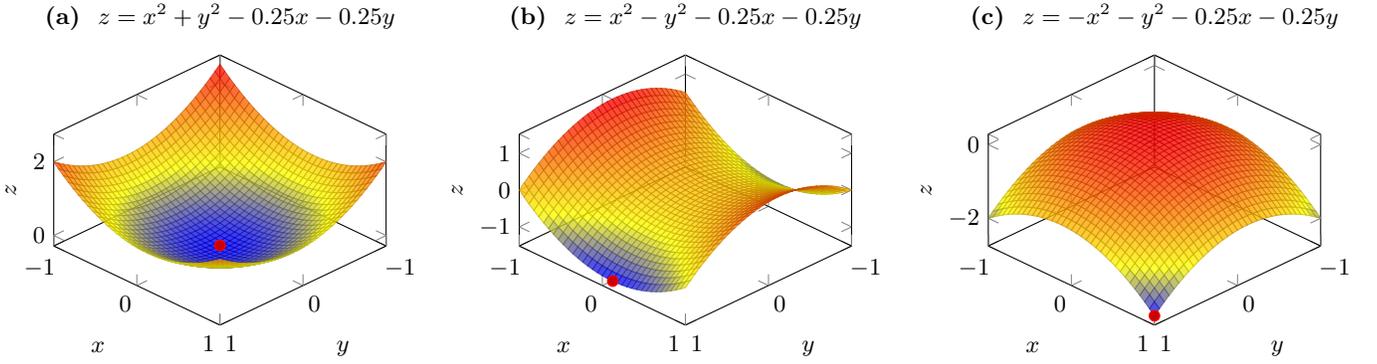

\begin{figure}[b]
  \includegraphics[width=0.75\linewidth]{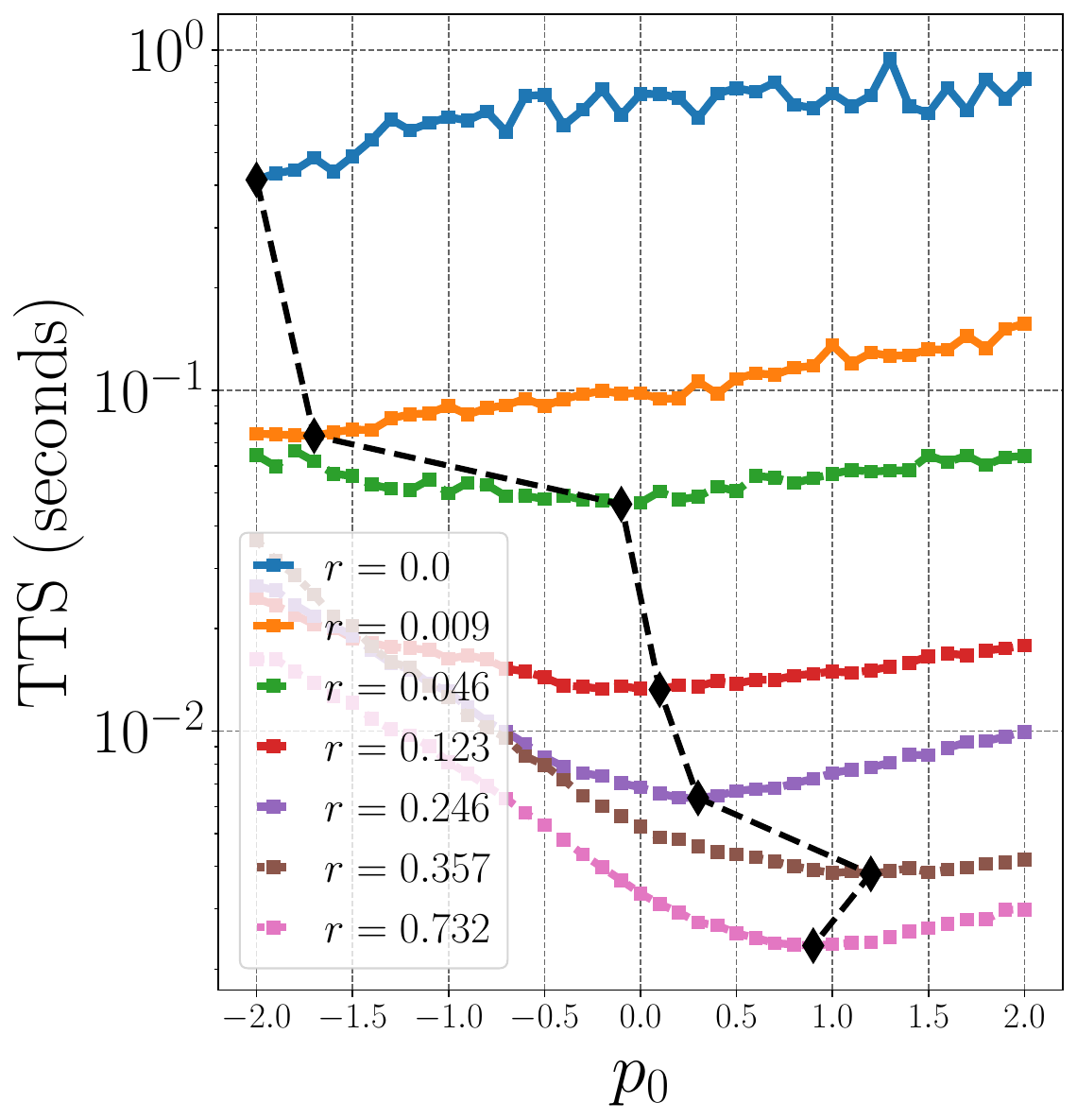}
  \caption{TTS versus pump value for random instances with different degrees of convexity solved using the MF-CIM dynamics. The ratio $r$ is here defined as the average number of fractional values in the solution vector of the generated problems instances divided by the problem size. An $r$ value of $0.0$ means that all the solutions are at the boundaries of the box while $r=1.0$ indicates a solution vector with only fractional values. Note that here the pump $p$ written in \cref{eq:sde_MF-CIM} is defined as $p=p_0(t/T) + (1+j) + g^2\tilde{\mu}_i^2$. This means that for $p_0=0$ in the figure, the effect of the extra drift term due to the nonlinear crystal is approximately cancelled out and the dynamics resemble those of the OLD. \label{fig:pump_vs_convexity}}
\end{figure}

As a practical application, we can leverage our observations to boost the performance of MF-CIM over the basic OLD depending on the geometry of the optimization landscape. A schematic of the various types of optimization landscapes is shown in \cref{fig:surfaces}. To this end, we vary the pump value of the MF-CIM when solving BoxQP problems as shown in \cref{fig:pump_vs_convexity}. The pump $p$ in \cref{eq:sde_MF-CIM} is written as $p = p_0(t/T) + (1+j) + g^2\tilde{\mu}_i^2$, yielding
\begin{equation}
  d\mu_i \approx  p_0\frac{t}{T} \mu_i\, dt - \lambda \partial_i f(\tilde\mu)\, dt + \sqrt{j}(\sigma_i - 1/2) dW_i,
\end{equation}
for the SDE of the mean field amplitudes in MF-CIM. Therefore, in \cref{fig:pump_vs_convexity} the TTS for different values of $p_0$ are shown. Each curve represents a group of randomly generated instances with varying degrees of convexity. The parameter $r$ is defined as the average number of fractional values in the solution vector of the problem instances divided by the problem size. A value of $r=1.0$ indicates that all components of the solutions are fractional, while a value of $r=0.0$ means that all variables attain their extreme values (binary solutions). The method for generating these instances is described in \cref{sec:app-instance_generation}.

As can be seen in \cref{fig:pump_vs_convexity}, for instances with the majority of solutions being binary ($r=0.0$ and $r=0.009$ instances), negative pump values of $-2.0$ and $-1.7$ (marked by black diamond points on the curves) provides the best TTS. This can be explained by noting that since majority of axes of these problem instances are concave with solutions at the boundaries, attracting the pulse amplitudes to the origin with a negative pump value encourages the solver to hill climbing over the concave landscape to improve the chance of escaping the local minima. On the other hand, for instances with higher numbers of fractional solutions ($r=0.357$ and $0.732$), a positive pump, which repels the pulse amplitudes away from the origin encourages a fast descent to the global minimum. These results provide intuitions on how to modulate the pump of the CIM to encourage exploration versus exploitation depending on the features of the optimization problem at hand. Moreover, interestingly, the ``more convex'' problem instances require fewer number of iterations and perform better with smaller or zero pump value, implying that they can be solved with lower energy consumption.

\section{Non-Evident Quantum Advantage\\(The Bad)}
\label{sec:the_bad}

\begin{figure*}[t]
  \subfloat[\label{fig:tts-FPGA-gaps}]{\includegraphics[width=0.285\linewidth]{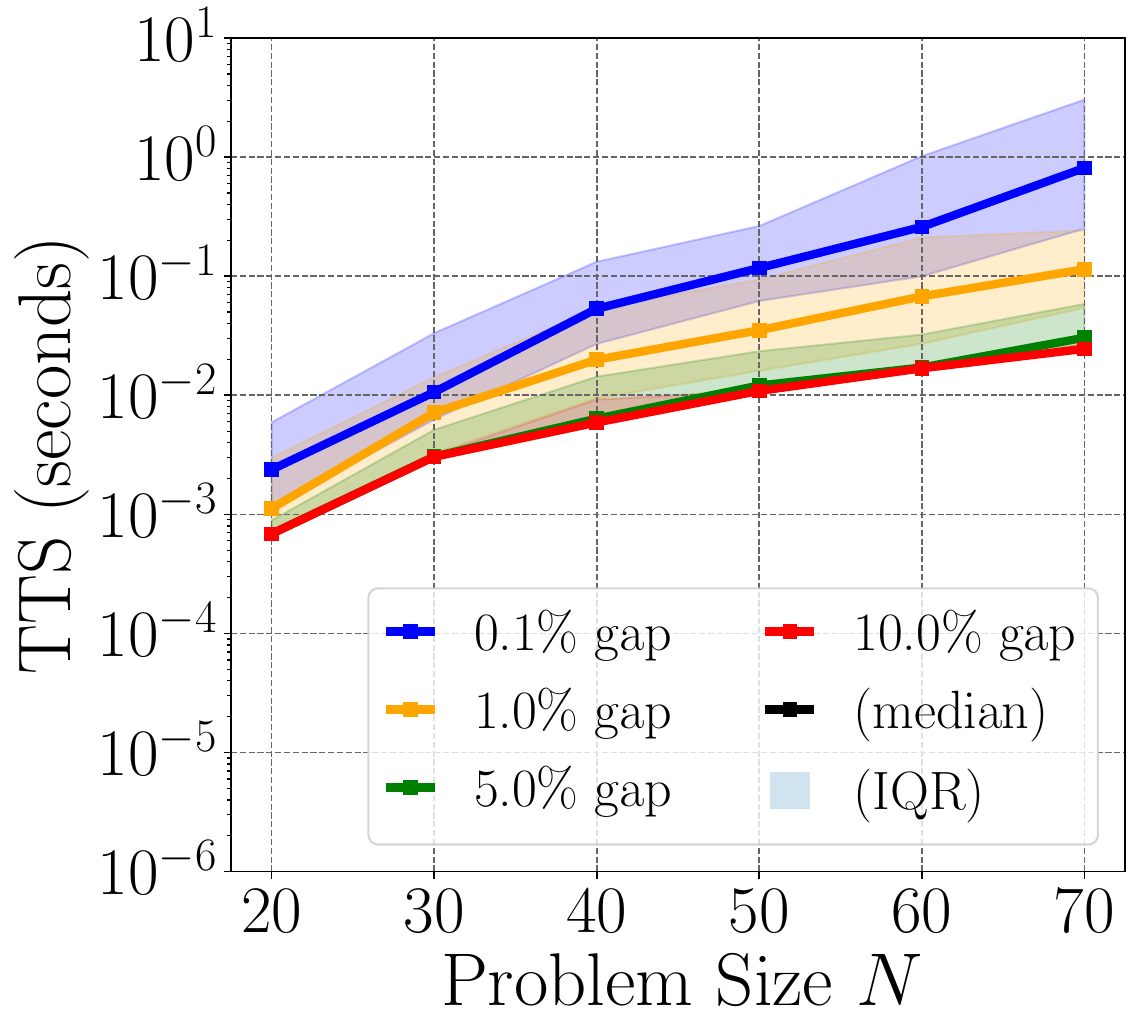}}\hspace{2mm}
  \subfloat[\label{fig:tts-MF-gaps}]{\includegraphics[width=0.285\linewidth]{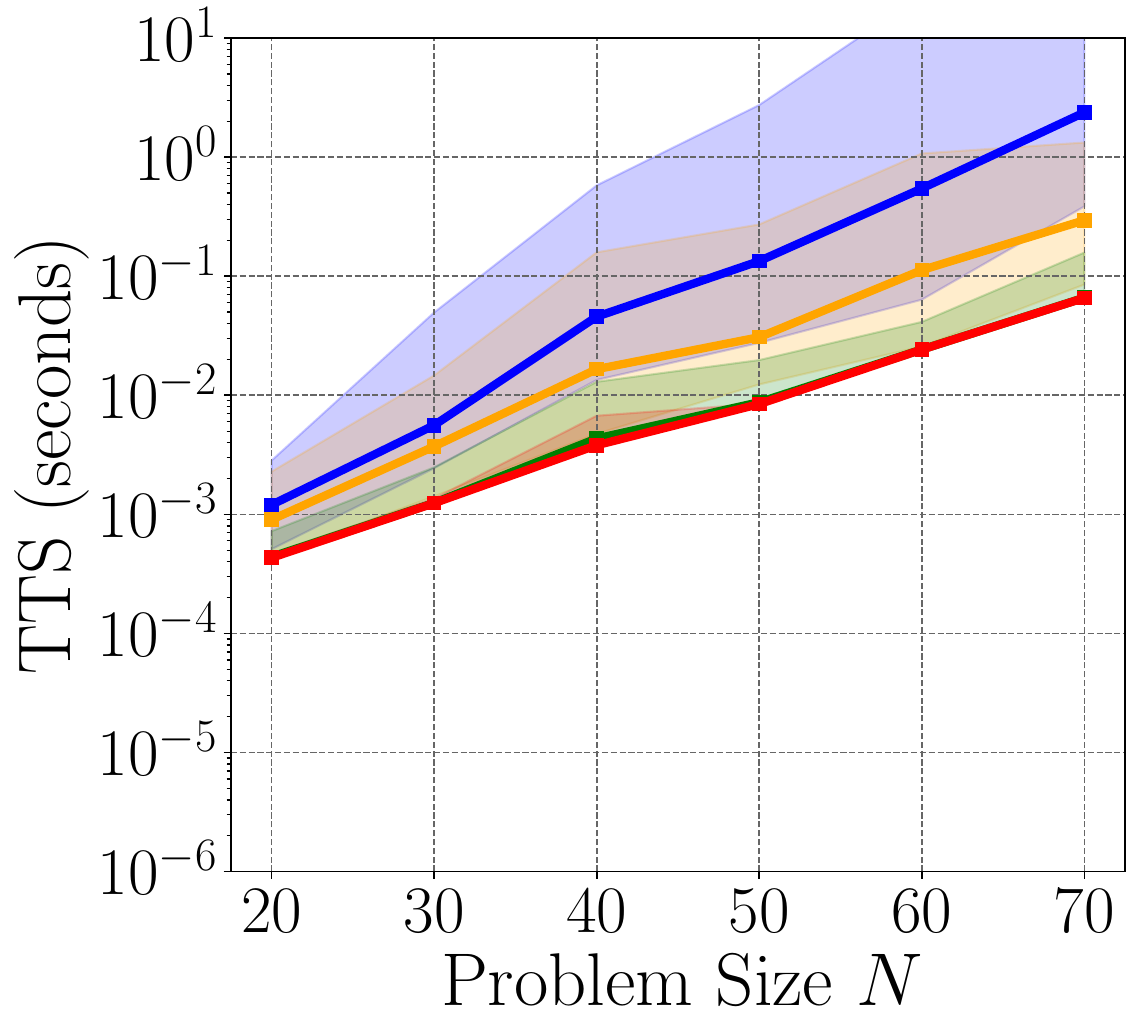}}
  \subfloat[\label{fig:tts-DL-gaps}]{\includegraphics[width=0.285\linewidth]{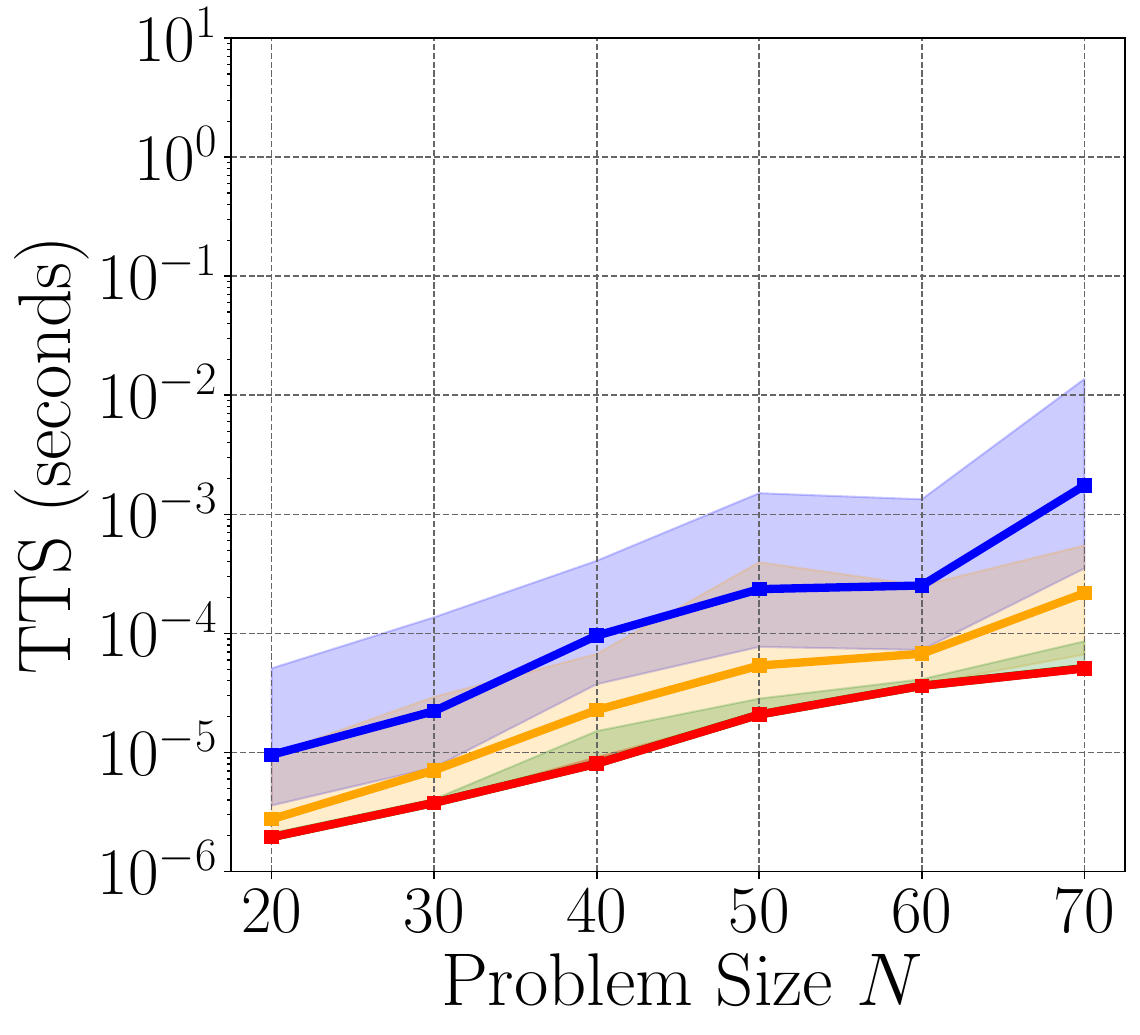}}
  \caption{Estimated TTS for (a) the CV-OLD solver implemented on FPGA, (b) the CV-MF-CIM, and (c) the CV-DL-CIM solvers. For each solver, several TTS curves are displayed for achieving various optimality gap targets. Solid curves show the TTS median values and the shaded regions show the IQR corresponding to the 25th and 75th percentile of the solved instances up to the specified optimality gap.
  \label{fig:tts-serial-gaps}}
\end{figure*}

As the results of the previous section suggest, although CIMs can offer improvements over simple OLD, they do not perform fundamentally more efficient than OLD when solving non-convex optimization problems. We now study the TTS scaling of our introduced solvers with respect to the problem size. \cref{fig:tts-serial-gaps} shows the TTS curves for BoxQP problems generated using the method of \cref{sec:app-instance_generation}.

\Cref{fig:tts-serial-gaps} displays the TTS median curves and interquartile range (IQR) regions for finding an approximate solution that is correct up to a multiplicative error (i.e., optimality gap). The OLD algorithm is implemented on an AMD Virtex UltraScale+ FPGA~\cite{FPGA_Virtex} with a clock speed of $300$ MHz (\cref{fig:tts-FPGA-gaps}). For the DL-CIM and MF-CIM, the physical optical device delays are estimated based on the experimentally motivated specifications detailed in \cref{sec:tts_appndx}.

We observe that DL-CIM (\cref{fig:tts-DL-gaps}) achieves about two orders of magnitude superior TTS than both the FPGA-based OLD (\cref{fig:tts-FPGA-gaps}) and the MF-CIM (\cref{fig:tts-MF-gaps}). The MF-CIM solver has a slightly worse overall TTS scaling compared to both OLD and DL-CIM with an exponential TTS growth law of slopes $0.043$--$0.063$ for optimality gap targets decreasing from $10\%$ to $0.1\%$. But DL-CIM with an exponential TTS growth law of slope $0.029$--$0.042$ is not statistically superior to the FPGA-based OLD with the slopes ranging in $0.029$--$0.049$. This, together with the correlation studies performed in \cref{sec:the_good} suggests that, while the fully optical DL-CIM takes advantage of the larger bandwidth in optics to perform computations faster than electronic and hybrid opto-electronic devices, it does not present a fundamental algorithmic advantage over the OLD as both attain similar asymptotic scaling laws with respect to the problem size.

While CIMs are theoretically shown to exhibit quantum effects, stringent conditions are placed on the parameters of the device (e.g. loss or coupling rate) for generating entangled states~\cite{zhou2021generating}. This severely limits the capability to perform quantum parallel search by using quantum entanglement or quantum superposition, the effects that are nevertheless destroyed through processes such as continuous measurement or photon loss across numerous delay line channels. It is therefore plausible to conclude that both DL-CIM and MF-CIM essentially perform approximate and slightly biased variants of the classical overdamped Langevin dynamics.

\section{No Leeway for Digital Shortcuts\\(The Ugly)}
\label{sec:the_ugly}

\begin{figure}[b]
  \subfloat[\label{fig:tts-serial}]{\includegraphics[scale=0.235]{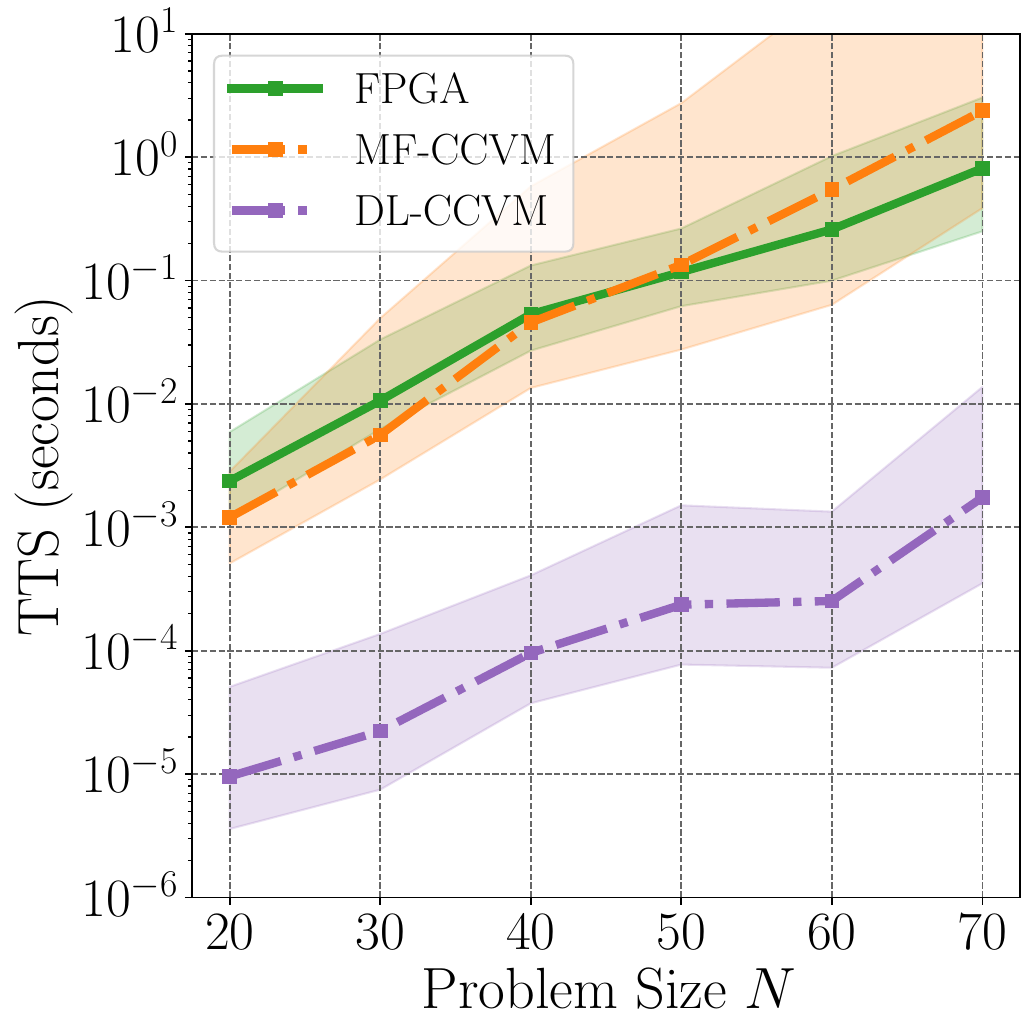}}\hspace{2mm}
  \subfloat[\label{fig:ets-serial}]{\includegraphics[scale=0.235]{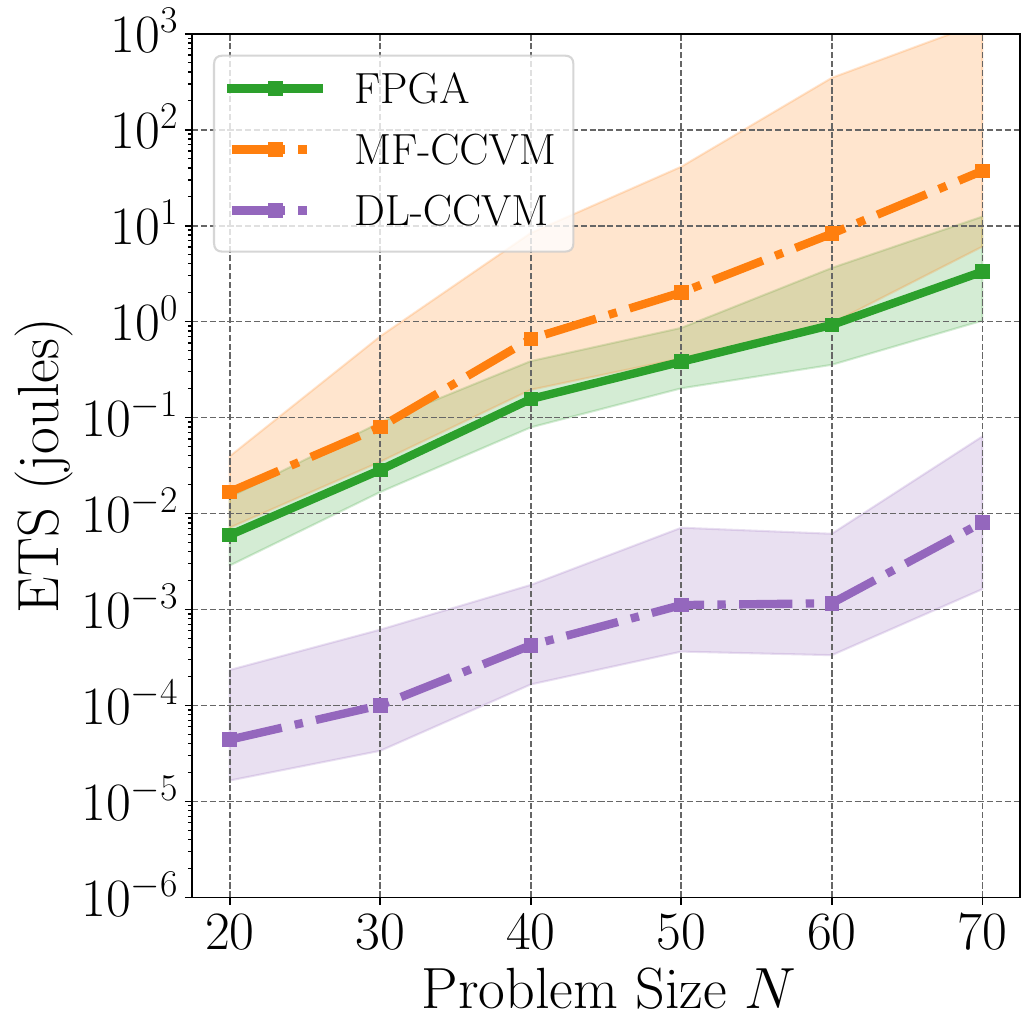}}
  \caption{(a) Estimated TTS and (b) ETS metrics for the three types of hardware studied here: FPGA-based OLD, CV-MF-CIM, and CV-DL-CIM. For each solver the median and the IQR range is presented and a common $0.1\%$ target optimality gap is assumed. The FPGAs in the full FPGA implementation (green curve) and the hybrid FPGA-optics implementation of MF-CIM (orange curve) are designed such that they have the smallest unit of vector-matrix multiplication. We call this implementation \textit{serial} and it provides a realistic performance curve for the FPGA and MF-CIM as the FPGA architecture is scalable for arbitrarily large problem sizes, assuming a scalable random access memory. We have assumed that the FPGAs in both of these designs to be from the AMD Kintex UltraScale+ family for the purposes of energy consumption calculations~\cite{FPGA_Virtex}.
  \label{fig:tts&ets-serial}}
\end{figure}

As previously shown in \cref{fig:tts-serial-gaps}, the FPGA-based implementation of OLD slightly outperforms MF-CIM in TTS scaling. Having assumed almost identical FPGA architectures for both solvers in our benchmarks suggests that the fast optics of MF-CIM (a pulse rate of $5$ GHz) is bottlenecked by the slow digital processor. In this section we further investigate this inferior performance in \cref{fig:tts&ets-serial}. As shown in \cref{fig:tts-serial}, the FPGA-based CV-OLD slightly outperforms MF-CIM in scaling, while DL-CIM achieves two orders of magnitude faster TTS as a prefactor.

The main differentiating factor between MF-CIM and FPGA-based OLD is the optical production of Wiener increments through continuous weak measurements, as opposed to generating pseudo-random number inside the FPGA. However, in practice digital pseudo-random number generation is not the most expensive subroutine of OLD. Moreover, the digital latency of pseudo-random number generation can be easily optimized by appropriate parallel processing within the FPGA (see \cref{sec:FPGA_parallelization}).

The drawbacks of the hybrid digital-optical scheme in MF-CIM is further evidenced by the ETS scaling results in \cref{fig:ets-serial}. Interestingly, the MF-CIM has an even larger energy consumption compared to an FPGA-only implementation of OLD. This is due to the presence of analog-to-digital (ADC) and digital-to-analog (DAC) convertors, as well as transceivers on the FPGA in the MF-CIM, none of which are needed in the FPGA-based OLD (see \cref{sec:energy-consumption}). Here again, the true merit of optical information processing is observed in the ETS scaling curve of the DL-CIM, revealing that fully optical computational devices can substantially reduce energy consumption beside the computation time.

For example, in the FPGA-only implementation of OLD, the total power consumed by the FPGA board for a problem of size $70$ is about $4.09$ W. About $6.5\%$ ($264$~mW) of this amount is used for random number generation, $43\%$ ($1.76$ W) for matrix-vector multiplication (MVM) and other mathematical operations, and the rest ($2.08$ W) is the static power to run the FPGA, board, and the clocking components. On the other hand, for the FPGA of the MF-CIM system, the total power consumed by the FPGA board is about $15.69$ W. From this amount, about $12.4\%$ ($1.94$ W) is consumed for MVM and other mathematical operations, $21.06\%$ ($3.3$ W) for the static power to run the FPGA, the board, and the clocking circuits, and the rest ($10.45$ W) to run the transceivers on the FPGA (for connectivity between the internal and external modules of the FPGA) as well as the DAC and ADC modules. Therefore the advantage of absence of digital random number generation in the MF-CIM device is completely lost due to the high power consumption of the DAC, ADC, and other transceiver operations required to convert between optical pulses and digital signals.

In this benchmark, both FPGAs used for the FPGA-based OLD and for the MF-CIM implementations use the smallest unit of MVM which is a single multiplication and addition at a time (see~\cref{sec:FPGA_parallelization}). This \textit{serial} implementation provides a smooth TTS and ETS scaling curves for comparing digital and optical hardware as the computational units and random access memory (RAM) can be assumed to increase proportionally with respect to problem sizes. While it is possible to improve the TTS and ETS metrics of the MF-CIM using parallelization and multiple FPGA hardware units~\cite{honjo2021spin}, the same tricks can be applied to the FPGA-based OLD implementation, again eliminating the chance of any significant advantage from optics in hybrid electronic-optical CIMs (see \cref{sec:FPGA_parallelization}). The advantage of optical random number generation indeed becomes more apparent when using a \textit{parallel} MVM implementation within the FPGA. In this case, the power consumed for random number generation is about $50\%$ of the total power consumption. Nevertheless, the hybrid system still incurs a large DAC and ADC power consumption, about twice as large as the power consumed for digital random number generation (see \cref{fig:fpga_power_division}). These observations show that analog optical computation is only truly beneficial when no DAC, digital processing, and ADC units are used, encouraging research towards fully analog and scalable optical information processing.

\section{\label{sec:conclusion} Conclusion}

We have presented the advantages and limitations of the coherent Ising machines (CIM) through a case study of the box-constrained quadratic programming (BoxQP) problem, the simplest NP-hard optimization problem that can be studied in the context of both binary and continuous optimization. We have shown that CIMs are intrinsically analog devices, that is, information is encoded in analog signals and processed by manipulating these signals continuously in time. These devices perform an approximate integration of the overdamped Langevin dynamics (OLD). This is true whether the problem solved is a binary-variable (BV) or a continuous-variable (CV) optimization problem. Viewing CIMs as analog CV optimizers allows us to avoid the overheads of digital (i.e., binary) information encoding, discretizations and problem reformulations in practical applications.

We have further shown that the CIMs do not present any fundamental computation advantage over OLD. While these devices are smartly quantum engineered to generate squeezed states of light, quantum computational effects such as entanglement and superposition are suppressed by continuous loss and measurement processes present in these devices. Additionally, we have argued that the hybrid optical-digital MF-CIM is limited by an analog-digital conversion bottleneck and therefore shows no significant advantage over digital-only devices such as FPGAs. In fact, we observed that MF-CIM consumed even more energy than an FPGA-only implementation of OLD. Our benchmarking study demonstrated that fully optical CIMs can solve non-convex continuous optimization problems around two orders of magnitude faster than classical and hybrid optical-digital heuristic solvers. Therefore we expect that the true merit of fast optics is in devices that avoid digital information conversions and operate fully in the analog regime.

Finally, we note that performant fully optical CIMs can have broader applications beyond quadratic optimization. Variants of these devices can be beneficial in all classical computational tasks that require solving stochastic differential equations (SDE). Generating high-quality random numbers and integrating SDEs at extremely high speeds and low energy consumption are the main advantages of optics in solving SDEs. The challenge, however, remains to be the implementation of efficient and reliable nonlinear interactions without resorting to digital electronics.

\section*{Data Availability}
The datasets generated and analyzed as part of our study
are available from the corresponding author upon
request. Our simulators of continuous variable CIM machines are publicly available as an open source Python package at \url{https://github.com/1QB-Information-Technologies/ccvm/}.

\section*{Acknowledgement}
The authors thank Yoshihisa Yamamoto, Edwin Ng, Satoshi Kako, Sam Reifenstein, Davide Ventureli, Robin Brown, and David Bernal for helpful discussions, as well as Mehmet Canturk for providing valuable feedback. We thank our editor, Marko Bucyk, for his careful review and editing of the manuscript. The authors acknowledge the financial support received through the NSF’s CIM Expeditions award (CCF-1918549). P.~R.~acknowledges the financial support of Mike and Ophelia Lazaridis, Innovation, Science and Economic Development Canada (ISED), and the Perimeter Institute for Theoretical Physics. Research at the Perimeter Institute is supported in part by the Government of Canada through ISED and by the Province of Ontario through the Ministry of Colleges and Universities.

\section*{Author Contributions}
Under the guidance of A.~S.\ and P.~R., F.~K. and M.~P.\ conducted the numerical experiments involved in this work. F.~K., A.~S., and P.~R. co-wrote the manuscript. All authors contributed to ideation throughout the research process. P.~R. led the overall efforts of this work.

\section*{Competing Interests}
The authors declare no competing interests.

\appendix
\section*{Appendix}

\section{Second derivative of Wiener processes and the MF-CIM dynamics}
\label{sec:app-distributions}

\Cref{eq:sde_linear-MF-CIM} for the Linear MF-CIM (LMF-CIM) involves the second derivative of the Wiener process, or the first derivative of white noise $\dot W= dW/dt$:
\begin{equation}
\begin{split}
d \mu_i =& -(1+j) \mu_i dt - \lambda \partial_i f( \mu)\, dt \\
&+ (1+j) \sqrt\frac{1}{4j} \dot W_i\, dt + \sqrt\frac{1}{4j} \ddot W_i\, dt\,.
\end{split}
\label{eq:sde_linear-MF-CIM-2}
\end{equation}
Unlike $\dot W$, the second derivative $\eta= \ddot W$ is ``too singular'' to be integrable as an It\^o process. Nevertheless, we can make sense of it using the theory of \emph{generalized} stochastic processes \cite{hida1980brownian, gel2014generalized} as briefly introduced below.

Given a probability space $(\Omega, \mathcal F, \mathbb P)$, one approach for defining a generalized process $\eta$ on it \emph{distributionally} is by determining what $\eta$ does on a family of time-dependent \emph{test functions}. Typically the set $\mathcal{D}= C_c^\infty(\mathbb R)$ of smooth compactly supported functions is used. This means associating an ordinary random variable
\begin{equation}
X_{\varphi} := \eta(\varphi) := \langle \eta, \varphi \rangle: \Omega \to \mathbb R
\label{eq:distribution-action}
\end{equation}
for any test function $\varphi$. If the mapping $\varphi \mapsto X_\varphi$ from $\mathcal D$ to the space of random variables $L^0 (\Omega)$ is linear and continuous, then $\eta$ is called a generalized stochastic process. Note that the bracket is merely a formal notation and not necessarily an inner product since $\eta$ is not generally an integrable function. However, the usual inner product $\langle \eta, \varphi \rangle = \int \eta\, d\varphi$ satisfies this definition when $\eta$ is integrable.

For example, white noise, $\xi= \dot W$, acts on every compactly supported test function $\phi$ via
\begin{equation}
X_{\varphi} = \langle \dot{W}, \varphi \rangle = - \langle W, \dot{\varphi} \rangle\,
\label{eq:white-noise-general-definition}
\end{equation}
using the usual $L^2$ inner products. Indeed, given that $W(t)$ is almost surely continuous and $\varphi$ is smooth with compact support, after integration by parts we have:
\begin{equation}
\int_{\mathbb R} \dot{W}(t)\, \varphi(t)\, dt
= \Big[ \varphi(t)\, W(t) \Big]_{-\infty}^\infty
- \int_{\mathbb R} W(t)\, \dot\varphi(t) dt.
\end{equation}
The boundary term vanishes away from the compact support and therefore,
\begin{equation}
X_\phi(\omega) = - \int_{\mathbb R} W(t, \omega)\, \dot{\varphi}(t)\, dt,
\end{equation}
for all $\omega \in \Omega$. We can indeed recover the usual properties of white noise. For example,
\begin{equation}
\mathbb{E}[X_{\varphi}(\omega)]
= -\int_{\mathbb R} \mathbb{E}[W(t,\omega)]\, \dot{\varphi}(t)\, dt = 0,
\end{equation}
and
\begin{equation}
\begin{split}
\mathbb{E}[X_{\varphi}(\omega) X_{\psi}(\omega)]
&= \iint \mathbb{E}[W(s,\omega) W(t,\omega)]\, \dot{\varphi}(s) \dot{\psi}(t)\, ds\, dt \\
&= \iint \min(s,t)\, \dot{\varphi}(s) \dot{\psi}(t)\, ds\, dt \\
&= \int_{\mathbb R} \varphi(t)\, \psi(t)\, dt,
\end{split}
\end{equation}
where the last equality follows from two integration by parts. In other words, the covariance operator $\mathcal C$ satisfying
\begin{equation}
\mathbb{E}[X_{\varphi} X_{\psi}] = \langle \mathcal{C} \varphi, \psi \rangle
\end{equation}
is identity $\mathcal C = I$. Since identity is identical to convolution with the $\delta$ function, the covariance kernel function of this process is $\delta(s-t)$.

Now, we can define the derivative of white noise, $\ddot{W}$, via two integration by parts as
\begin{equation}
\langle \ddot{W}, \varphi \rangle := \langle W, \ddot{\varphi} \rangle.
\label{eq:derivative-white-noise-general-definition}
\end{equation}
Note that similar to $\dot W$, the second derivative $\ddot{W}$ is also a (generalized) Gaussian process. A generalized Gaussian random process $\eta$ is one for which any random vector formed by testing against $N$ functions is (multivariate) Gaussian, that is, if
\begin{equation}
X_{\varphi_1:\varphi_N} = [\langle \eta, \varphi_1 \rangle,\, \ldots,\, \langle \eta, \varphi_N \rangle]^T \in \mathbb{R}^N,
\end{equation}
is Gaussian for every choice of test functions $\varphi_i$. Then, Gaussianity follows from the definitions in \cref{eq:white-noise-general-definition,eq:derivative-white-noise-general-definition}.

The mean of $\ddot W$ is also zero since $\langle \mathbb{E}[\ddot{W}], \varphi \rangle = \langle \mathbb{E}[W], \ddot{\varphi} \rangle = 0$. However, $\ddot W$ is still a different process. For example, the covariance kernel is not the same as $\delta(s - t)$. We have
\begin{equation}
\begin{split}
\mathbb{E}[X_{\varphi} X_{\psi}]
&= \iint_{-\infty}^{\infty} \min(s,t)\, \ddot{\varphi}(s) \ddot{\psi}(t)\, ds\, dt \\
&= -\int_{-\infty}^{\infty} \ddot{\varphi}(t)\, \psi(t)\, dt
= \langle \mathcal{C} \varphi, \psi \rangle.
\end{split}
\end{equation}
Thus the covariance operator is $-d^2/ dt^2$, and the covariance kernel function is $-\delta''(s-t)$.

Returning to the LMF-CIM equation \cref{eq:sde_linear-MF-CIM-2}, we can make sense of this stochastic process, by observing its effect on any test function $\varphi(t)$ as follows:
\begin{equation}
\begin{split}
\int_{-\infty}^\infty \varphi(t)\, \dot \mu_i (t)\, dt
&= \int_{-\infty}^\infty \varphi(t)
\Big(
- (1 + j)\, \mu_i
- \lambda\, \partial_i f(\tilde{\mu})
\Big) dt \\
&+ (1 + j)\, \sqrt{\frac{1}{4j}}\, \int_{-\infty}^\infty \varphi(t)
\dot W (t)\, dt \\
&+ \sqrt{\frac{1}{4j}}\, \int_{-\infty}^\infty \varphi(t)
\ddot W (t)\, dt
\end{split}
\end{equation}
Integrating by parts for the derivative terms yields
\begin{equation}
\begin{split}
\int_{-\infty}^\infty \mu_i (t)\, \dot \varphi(t) \, dt
&= \int_{-\infty}^\infty \varphi(t)
\Big(
(1 + j)\, \mu_i
+ \lambda\, \partial_i f(\tilde{\mu})
\Big) dt \\
&+ (1 + j)\, \sqrt{\frac{1}{4j}}\, \int_{-\infty}^\infty
W(t) \dot \varphi(t)\, dt \\
&- \sqrt{\frac{1}{4j}}\, \int_{-\infty}^\infty
W (t) \ddot \varphi(t)\, dt.
\end{split}
\end{equation}
This distributional identity must hold for all test functions $\varphi \in C_c^\infty(\mathbb{R})$.

\section{Solving B\lowercase{ox}QP problems using CIMs}
\label{sec:solvingbyCIM}

In this section, we follow the framework introduced in \cite{khosravi2022non} for solving continuous variable problems using the CIM. We introduce the box-constrained quadratic problems (BoxQP) problem, our approach for implementing and solving them using CIMs, and a new method for generating random BoxQP instances.

\subsection{Box-constrained quadratic programming (BoxQP) problems}

The BoxQP problem can be formulated as follows:
\begin{equation}
\begin{split}
\text{maximize} \quad& f(x) = \frac{1}{2} \sum_{i,j=1}^N Q_{ij} x_i x_j + \sum_{i=1}^N V_i x_i  ,\\
\text{subject to}\quad & \ell_i \le x_i \le u_i \quad \forall i \in \{1, \ldots, N\},
\end{split}
\label{eq:boxQP}
\end{equation}
where $Q\in\mathbb R^{N \times N}$ is a symmetric matrix, $V\in\mathbb{R}^N$ is a real $N$-dimensional vector, and the lower and upper bounds $\ell_i\in\mathbb{R}$ and $u_i\in\mathbb{R}$ specify the box constraints. Here, for simplicity, we will assume all $\ell_i= 0$ and $u_i= 1$ and refer the reader to \cite{khosravi2022non} for a thorough study of these problems.
In this paper, e.g., in \cref{fig:correlations}, we have solved both binary-variable (BV) and continuous-variable (CV) versions of the BoxQP problem. As shown above, the BoxQP problem is originally a CV problem. To convert it to a BV problem, we replace the box constraints with binary variable constraints, $x_i\in \{0, 1\}$ for all $i \in \{1,\cdots,N\}$. This restriction may or may not change the optimal solution of the problem (see \cref{fig:pump_vs_convexity}).

\subsection{Mapping BoxQP problems on the CIM}

CIMs can be operated below the saturation threshold to solve CV optimization problems. In this case, the amplitude and phase of the pulses represent the value and the sign of the variables of the problem. By properly pumping the optical pulses, the CV variables of the BoxQP problem are encoded into the analog pulse amplitudes natively. Furthermore, the box constraint is either implemented using a digital processor, as in the case of the MF-CIM, or by exploiting the saturation feature of the nonlinear crystal in the case of the DL-CIM \cite{khosravi2022non}. 

Specifically, in the DL-CIM scheme, the following change of variable is performed, 
\begin{equation}
x_i \mapsto \frac{1}{2} \left(\frac{c_i}{s} + 1 \right),
\label{eq:x_subs}
\end{equation}
where $s\simeq\sqrt{p_0 - 1}$ is the approximate saturation amplitude for all the DOPO pulses. Here $p_0$ is the pump value at the end of the evolution process. Therefore, the drift term in the SDEs of the DL-CIM described by~\cref{eq:sde_DL-CIM} becomes~\cite{khosravi2022non}:
\begin{align}
  -\partial_i f(c)
 = & - \sum_{j=1}^{N}
\frac{Q_{ij}}{2s}
\left[\frac{1}{2}\left(\frac{c_j}{s} + 1\right) \right]
- \frac{V_i}{2s}.
\label{eq:DLcoupling}
\end{align}

In the MF-CIM scheme, we use a similar encoding to that of the DL-CIM,
\begin{equation}
x_i \mapsto \frac{1}{2}\left(\frac{\tilde{\mu_i}}{s} + 1\right),
\end{equation}
to map the problem, which results in the drift term
\begin{align}
-\partial_i f(\tilde{\mu})
 = & -\sum_{k=1}^{N}
\frac{Q_{ik}}{2s}
\left[\frac{1}{2}\left(\frac{\tilde{\mu}_k}{s} + 1\right)\right]
- \frac{V_i}{2s}.
\label{eq:MFcoupling}
\end{align}
Here, $s$ is a hyperparameter that represents a saturation bound~\cite{khosravi2022non}. As long as \mbox{$s < \sqrt{p - (1+j)}/ g$} near the end of the evolution, it is tuned to obtain the best performance.

\subsection{Randomly generated BoxQP instances}
\label{sec:app-instance_generation}

We have tested the performance of the various solvers introduced in this paper on randomly generated BoxQP problem instances. As opposed to the older methods~\cite{vandenbussche2005branch,khosravi2022non}, our new approach provides us control over the hardness of the problem instances and the number of fractional values in the corresponding optimal solutions. For simplicity, all variables are assumed to be constrained to the domain $[0,1]$ by setting $\ell_i = 0$ and $u_i = 1$ for all $i\in \{1,\ldots,N\}$. The global maxima of the generated instances were found using Gurobi~9.5~\cite{gurobi}. \cref{sec:time-evolution} further demonstrates the time-evolution of the dynamics of the DL-CIM and MF-CIM solvers for two example BoxQP problem instances.

The previous problem generation method used in Refs. \cite{vandenbussche2005branch,khosravi2022non} relies on constructing random $Q$ matrices and $V$ vectors whose elements are sampled from a symmetric distribution centered around zero. The disadvantage of this method is that the solution vector for the instances contains values that are mostly at the boundaries of the box constraints. In other words, there are very few fractional values between $0$ and $1$ in the optimal solutions. Since only the negative eigenvalues of $Q$ (in a maximization problem) can give rise to fractional solutions, the ratio between the elements of $V$ and the negative eigenvalues of $Q$ determine whether the solution along a given axis is fractional. With this ratio being itself a random number, the majority of the optimal values are at the boundaries of the box constraints. In fact, many of the instances generated using this method do not contain any fractional optimal values.

To mitigate this issue, we consider biasing the eigenvalues of the $Q$ matrix to contain more positive values. While, with proper setting of the range of values for the $V$ vectors, this can lead to more fractional values in the solution, this would lead to easier problem instances. This is because majorities of the axes of the generated instance are convex and a simple gradient descent approach can find most of the values of the solution vector.

To maintain the difficulty of the instances, we generate a random diagonal matrix $D$ representing the eigenvalues of the $Q$ matrix, and a random $C$ vector representing the $V$ vector before performing a rotation, with elements sampled from a uniform distribution between $-50$ and $+50$ for both $D$ and $C$. We then generate random $N\times N$ rotation matrices $R$ by performing 2D rotations repeatedly on randomly selected pairs of axes of an identity matrix until all the elements of the rotation matrix $R$ are non-zero. We then use this rotation matrix to produce the $Q$ matrix and the $V$ vector according to $Q=R\cdot D\cdot R^\text{T}$ and $V=R\cdot C$. Since the box constraint is still imposed in the new coordinates, the solution of the instance becomes intractable through the rotation process. The maximum angle for the random 2D rotations $\phi_\text{max}$ is a determining factor for the characteristics of the randomly generated instances. For small $\phi_\text{max}$, there are more fractional values in the solution but the instances are easier to solve as they are closer to the original diagonal problem. Larger values of $\phi_\text{max}$, however, lead to harder problem instances but with less fractional values. In this paper, we have set $\phi_\text{max}=\pi/2$ for generating the random instances used to evaluate the Langevin and CIM solvers. For each problem size, we have generated 50 random problem instances.

For generating the random instances with different degree of convexity in \cref{fig:pump_vs_convexity}, the random eigenvalues of the diagonal $D$ matrix were sampled from a uniform distribution from the range $-50+c_0$ and $+50+c_0$, where $c_0>0$ offsets the eigenvalues towards positive values, making the instances more concave (in a maximization problem), while $c_0<0$ pushes the eigenvalues towards negative values, making the problems convex along more of the axes. The factor $r$ in \cref{fig:pump_vs_convexity} is the ratio of fractional values in the solution of the problem instances to the number of variables, averaged over all the instances. Values of $c_0 = +50, +25, +12.5, 0, -12.5, -25, -50$ have led to $r$ values of $0.0$, $0.009$, $0.046$, $0.123$, $0.246$, $0.357$, and $0.732$, respectively, in \cref{fig:pump_vs_convexity}. For $r=+50$, for example, all eigenvalues of the $Q$ matrix are positive and thus all the solutions lie on the boundary of the box when maximizing the objective function constrained to the box. We also have set $\phi_\text{max}=0.1\pi$ for the instances used to generate the results in \cref{fig:pump_vs_convexity}.

\begin{figure}[!t]
\includegraphics[width=0.75\linewidth]{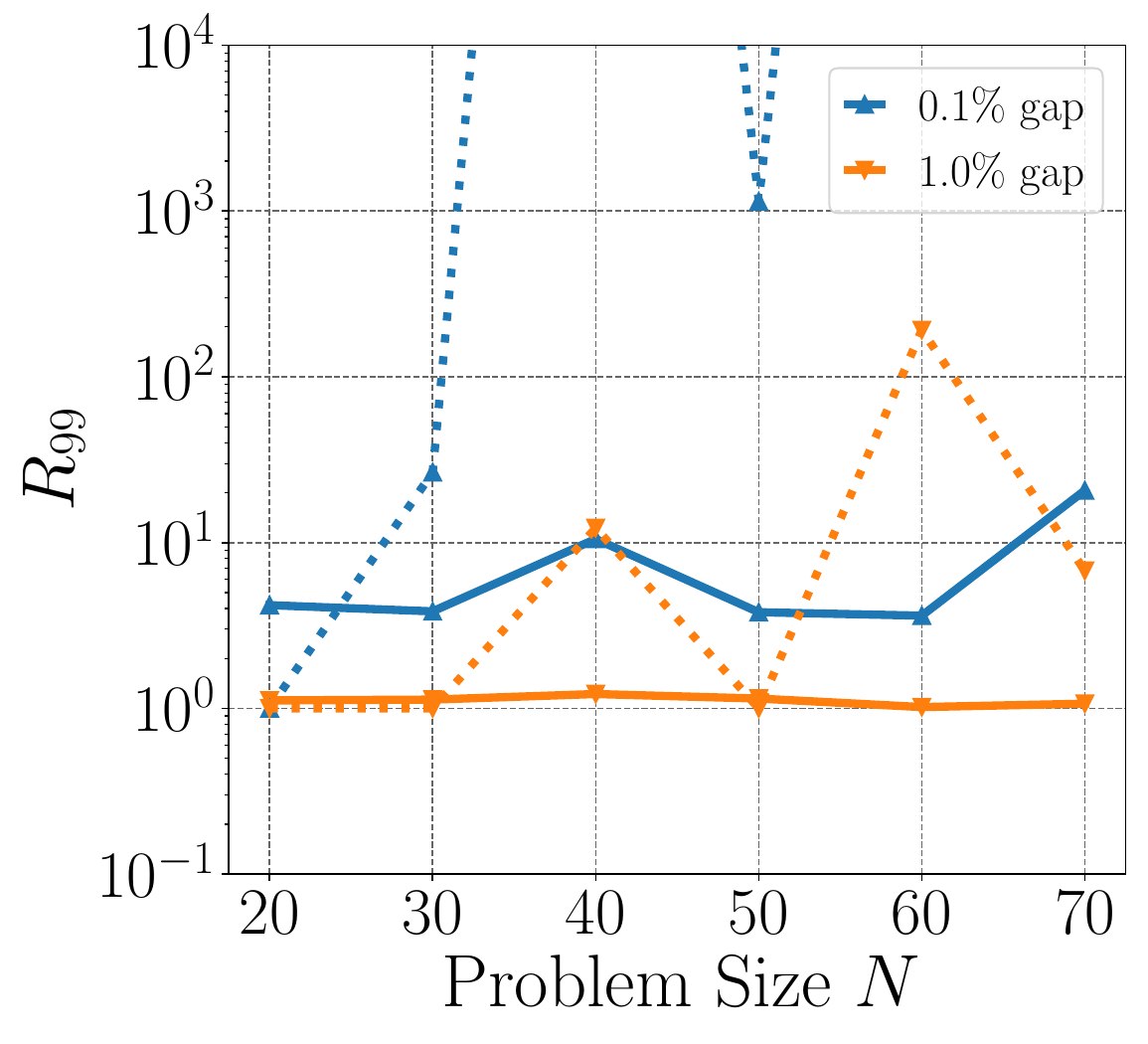}
\caption{The effect of the noise in the DL-CIM method (solid lines) as compared to solving the instances using a simple Euler gradient descent method with constraint enforcement at each step (dotted lines), when solving the generated random BoxQP instances. The smaller values for $R_{99}$ of the DL-CIM indicates that the noise in the process helps in escaping local minima and finding the optimum solutions of the problem instances. The $R_{99}$ value for the gap percentages of equal to or larger than $5.0\%$ is equal to $1.0$ for both methods and all instance sizes. \label{fig:noise_effect}}
\end{figure}

\subsection{Significance of CIM's noise injection\label{sec:effect_of_noise}}

\cref{fig:noise_effect} shows the effectiveness of the noise when solving the randomly generated instances. We have compared the performance of the DL-CIM method (solid lines) against a simple Euler gradient descent method (with no noise) with constraint enforcement at each step. For the $0.1\%$ gap (blue curves), the Euler gradient descent method fails to find a solution for more than $50\%$ of the instances for problem sizes $40$, $60$, and $70$, and it under-performs compared to the DL-CIM method for problem sizes $30$ and $50$. These results show that, due to the presence of local minima in the landscape of the instances, the noise helps the solver to escape the local minima and find the optimum solution. Similar observations are made for the other solvers studied in this paper.

\section{Benchmarking metrics\label{sec:benchmarking_metrics}}

\subsection{Time-to-solution calculations}
\label{sec:tts_appndx}

Here, we provide more details for estimating the TTS of the optical devices with the dynamics introduced in \cref{sec:langevin_and_con}. TTS is a useful metric for comparing the performance of different hardware when solving a given optimization problem \cite{khosravi2022non}. \cref{tab:solvers_parameters} shows the parameters of the solvers used for generating the results in \cref{fig:tts&ets-serial,fig:tts&ets-parallel,fig:tts-serial-gaps}. For the OLD, PDL, and DL-CIM solvers, the pump field $p(t) =\frac{t}{T_\text{max}} p_0$ follows a linear schedule, while for the MF-CIM it was set to \mbox{$p(t) = \frac{t}{T_\text{max}} p_0 + 1 + j(t)$} to compensate for the measurement loss and background loss.
On the other hand, the measurement strength is $j(t) = j_0 \exp\left(-\alpha \frac{t}{T_\text{max}}\right)$, where $\alpha$ is an arbitrary parameter. For the DL-CIM solver, the injected noise parameter is \mbox{$r(t) = r_0 \exp\left(-\beta \frac{t}{T_\text{max}}\right)$,} where $\beta$ is an arbitrary parameter~\cite{khosravi2022non}.

\begin{table*}[!ht]
\begin{tabular}{|c|c||c|c|c|c|}\hline
    Parameter & Variable & OLD & PLD & DL-CIM & MF-CIM \\ \hline
Number of Iterations & $n_\text{iter}$ & $500-1500$  & $500-1500$  & $10^3-10^4$ & $300-4000$ \\\hline
Laser Pulse Rate & $f_\text{laser}$ & - & - & $100$ GHz & $5$ GHz \\\hline
Normalized Pump Field & $p_0$ & - & $2.0$ & $8.0$ & $0.0$ \\\hline
Normalized Time Increment & $dt$ & $0.002$ & $0.002$ & $0.001$ & $0.0025$ \\\hline
Diffusion Parameter & $\sigma$  & $0.5$ & $0.5$ & - & - \\\hline
Injected Noise Parameters & $(r_0,\beta)$ & - & - & $(10,3)$ & - \\\hline
Normalized Optical Non-linearity in DL-CIM & $A_s$ & - & - & $10$ & -\\\hline
Saturation Parameter & $s$ & - & - & $1.2\sqrt{p_0-1}$ & $20.0$ \\\hline
Measurement Strength Parameters & $(j_0, \alpha)$ & - & - & - & $(5,3)$\\\hline
Gradient Function Strength & $\lambda$ & $1.0$ & $1.0$ & $50 + \frac{t}{T_\text{max}}100$ & $4000$ \\\hline
Normalized Optical Non-linearity in MF-CIM &  $g$ & - & - & - & $0.001$\\\hline
\end{tabular}
\caption{Parameters for the OLD, PLD, DL-CIM, and MF-CIM solvers.}
\label{tab:solvers_parameters}
\end{table*}

For the DL-CIM solver, the time for a single trial is estimated as  $T_\text{max} = n_\text{iter} \cdot N \cdot T_\text{pulse}$, where $T_\text{pulse}=1/f_\text{laser}$ is the time delay between two individual pulses within the cavity and $f_\text{laser}$ is the pulse rate of the laser source; for our benchmarking study, we have used the value \mbox{$f_\text{laser} = 100$ GHz} or $T_\text{pulse} = 10$ picoseconds~\cite{reifenstein2021coherent}. While in practice the size of the device's optical cavity is typically fixed, for our benchmarking study we have assumed that the length of the optical cavity is optimally adjusted to fit the exact number of the optical pulses representing the number of variables in a given problem instance in order to obtain the optimal TTS. In the case of the MF-CIM solver, the single trial time is estimated as $T_\text{max} = n_\text{iter}(T_\text{FPGA} + T_\text{buffer})$. Here $T_\text{FPGA}$ is the time required by the FPGA to compute the feedback term at every round trip. This time is accounted for in the optical cavity by increasing the length of the coil in between the two couplers depicted in~\cref{fig:ccvm-schematic}. The term $T_\text{buffer}$ is a small additional latency needed for the pulses to be amplified by the OPA before their arrival at the output coupler prior to the next round trip. In the MF-CIM, $f_\text{laser} = 5~\text{GHz}$ to match with the speed of commercially available DACs and ADCs~\cite{ADC_DAC}. The number of iterations used for solving each problem size using the DL-CIM and MF-CIM methods were found by tuning $n_\text{iter}$ (\cref{sec:tts_vs_iter}).

\cref{tab:FPGA_runtime} shows the latencies of the FPGA and its supporting components for each problem size studied. For the OLD solver, which is implemented on the FPGA using a basic arithmetic approach based on the Euler method, the FPGA time for each iteration is measured using the available FPGA metrics. For the FPGA implemented as part of the MF-CIM solver, the FPGA times are estimated by multiplying the total number of clock cycles required for evaluating the feedback term (for ADC, DAC, and the FPGA computations) by the clock duration at a clock frequency of $300$~MHz.

\begin{table}[!h]
\begin{tabular}{|c|c|c|c|c|c|c|}
\hline
Problem Size & $20$ & $30$ & $40$ & $50$ & $60$ & $70$ \\
\hhline{|=|=|=|=|=|=|=|}
OLD on FPGA & \multirow{2}{*}{$1.37$} & \multirow{2}{*}{$3.04$} & \multirow{2}{*}{$5.37$} & \multirow{2}{*}{$8.37$} & \multirow{2}{*}{$12.03$} & \multirow{2}{*}{$16.37$}\\
\textit{serial} ($\mu\text{s}$) & & & & & & \\
\hline
OLD on FPGA & \multirow{2}{*}{$0.053$} & \multirow{2}{*}{$0.053$} & \multirow{2}{*}{$0.057$} & \multirow{2}{*}{$0.057$} & \multirow{2}{*}{$0.057$} & \multirow{2}{*}{$0.06$}\\
\textit{parallel} ($\mu\text{s}$) & & & & & & \\
\hline
FPGA in MF-CCVM & \multirow{2}{*}{$1.43$} & \multirow{2}{*}{$3.1$} & \multirow{2}{*}{$5.43$} & \multirow{2}{*}{$8.43$} & \multirow{2}{*}{$12.1$} & \multirow{2}{*}{$16.43$ } \\
\textit{serial} ($\mu\text{s}$) & & & & & & \\
\hline
FGPA in MF-CCVM & \multirow{2}{*}{$0.093$} & \multirow{2}{*}{$0.093$} & \multirow{2}{*}{$0.097$} & \multirow{2}{*}{$0.097$} & \multirow{2}{*}{$0.097$} & \multirow{2}{*}{$0.1$ } \\
\textit{parallel} ($\mu\text{s}$) & & & & & & \\
\hline
\end{tabular}
\caption{Latency of the FPGA when implementing the OLD (first two rows) and the latency of the FPGA together with that of the ADC and DAC in the MF-CIM (second two rows) for the \textit{serial} and \textit{parallel} implementations. These values are for a single Euler method iteration in the FPGA implementation or a single round-trip time in the MF-CIM.}
\label{tab:FPGA_runtime}
\end{table}

Since the solutions of the problem instances are continuous numbers, the proposed stochastic solvers typically yield approximate solutions in close proximity to the optimum solutions due to the presence of noise. To address this issue, we performed a short post-processing step for all solvers where the solution found by the solver was used as the initial value for a simple Euler gradient descent method with constraint enforcement at each step. For each solver, the number of iterations chosen for this post-processing step was $1\%$ of the total number of iterations used by the solver in solving a given problem instance.

\subsection{Energy consumption estimates}
\label{sec:energy-consumption}
In this section, we present our approach for estimating the ETS for three solvers implemented on five different hardware: OLD implemented on FPGA, CPU, and GPU; DL-CIM; and MF-CIM. ETS is a metric that quantifies the energy consumption of a given hardware device in implementing a specified solver for a given BoxQP instance. Similarly to the way we define the TTS, we define $\text{ETS} = R_{99} \cdot E_\text{max}$, where $R_{99}$ is the number of trials needed to solve a given problem instance with a $99\%$ success rate at least once and $E_\text{max}$ is the estimated maximum energy consumption for a single run of a given solver. Here, $E_\text{max} = P_\text{max}\cdot T_\text{max}$, where $P_\text{max}$ is the power consumed by the entire device. Note that here $P_\text{max}$ may have contributions from different elements in a given device which may not be running continuously all at the same time. While this is a valid assumption, we ignore such considerations in order to find an upper limit on the energy consumption of a device.

\begin{figure}[!t]
  \centering
  \subfloat[\label{fig:fpga_powers_serial}]{\includegraphics[width=0.95\linewidth]{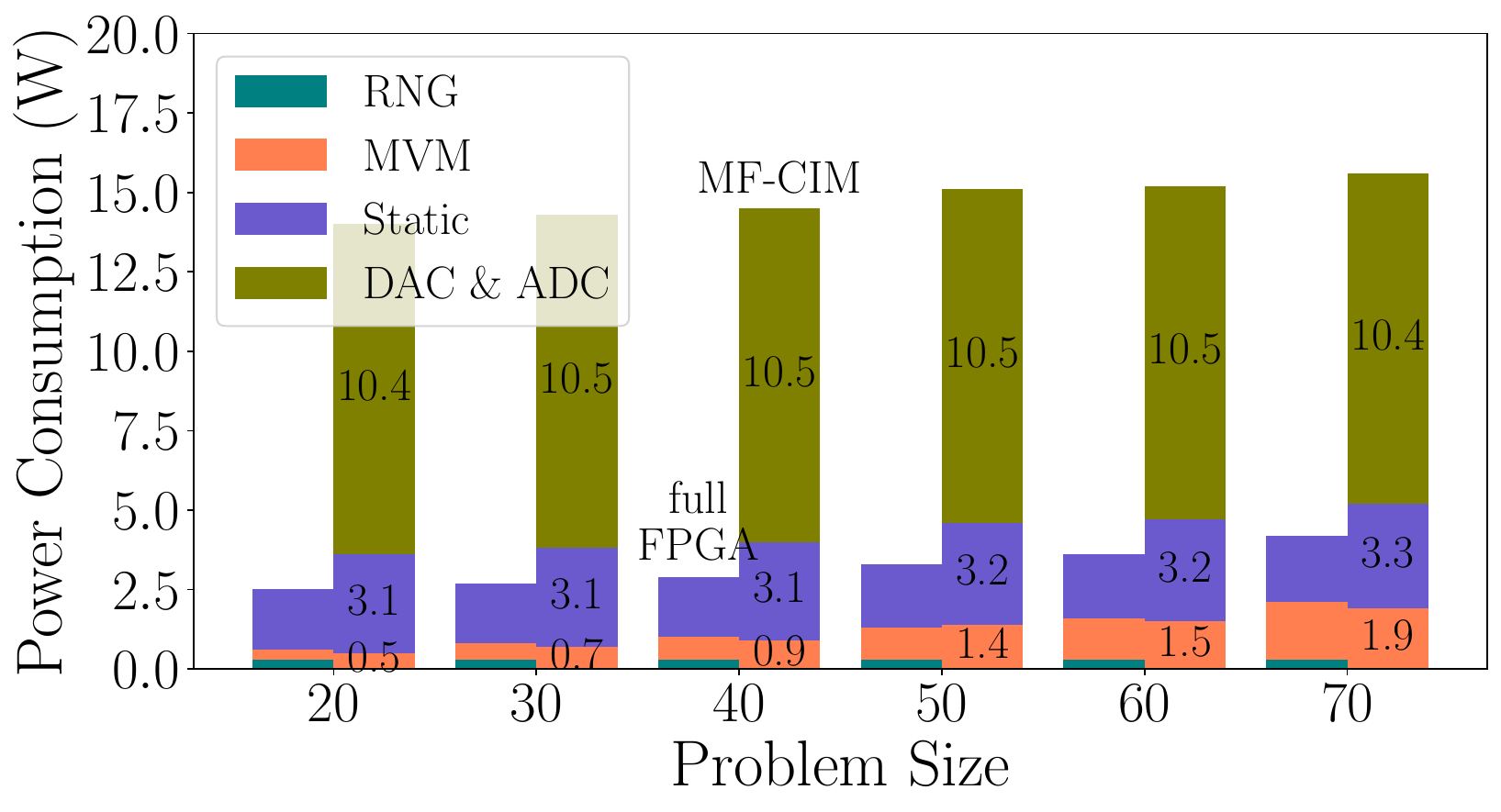}}\\
  \subfloat[\label{fig:fpga_powers_parallel}]{\includegraphics[width=0.95\linewidth]{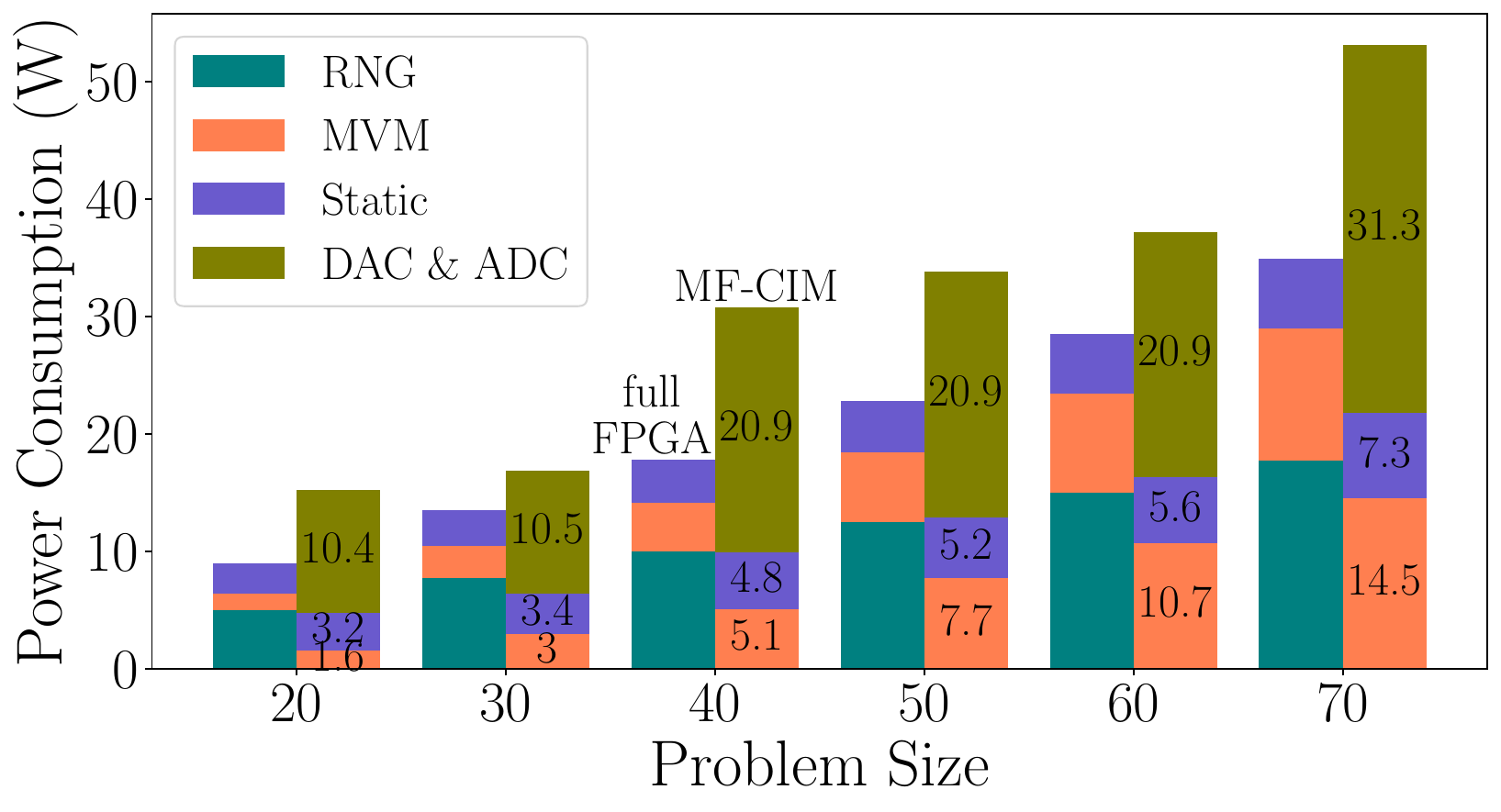}}
  \caption{The power consumption of the FPGA and the surrounding circuits for a single iteration of the Euler method in the OLD implemented on FPGA (``full FPGA") as well as the single calculation of the feedback term in MF-CIM (``MF-CIM"). (a) \textit{Serial} implementation based on \cref{fig:fpga_serial} and (b) \textit{parallel} implementation based on \cref{fig:fpga_parallel}. In the full FPGA implementation, the power is divided between RNG, MVM, and static power, while in the MF-CIM implementation the power is divided between MVM, DAC \& ADC, and static power. \label{fig:fpga_power_division}}
\end{figure}

\cref{fig:fpga_power_division} shows the power consumption of the FPGA and its supporting circuitry for the OLD implemented on the FPGA (``full FPGA'') and the MF-CIM (``MF-CIM'') for different problem sizes and for the \textit{serial} (\cref{fig:fpga_powers_serial}) and \textit{parallel} (\cref{fig:fpga_powers_parallel}) implementations (see \cref{sec:results,sec:FPGA_parallelization}). These values are obtained using metrics available for the FGPA. The total energy consumption of the FPGA for a single run of the OLD solver can then be found by multiplying these power consumption values by the latencies introduced in \cref{tab:FPGA_runtime}. Similarly, in the case of the OLD implemented on a GPU, the power consumption can be obtained from the metrics available by the card for an Nvidia T4 GPU~\cite{TeslaT4}. We measured a power consumption ranging from $28.93$ watts to $32.28$ watts for various problem sizes, with an idling power of about $9.96$ watts. In the case of the OLD solved on a CPU, the power consumption is estimated for the dynamics implemented on MacBook Pro laptop with an Apple M1 chip~\cite{AppleM1}.

For the MF-CIM device, the total power consumption can be written as $P_\text{max} = P_\text{opt} + P_\text{FPGA}(N)$, where $P_\text{opt}$ is the power required to generate the optical pulses, while $P_\text{FPGA}(N)$ is the power required by the FPGA to generate the feedback terms for a given problem size $N$. Taking into account the power required for second-harmonic generation and parametric down-conversion in a coherent optical network~\cite{mcmahon2016fully}, we take $P_\text{opt} \approx 1$ mW. To find the energy consumption, this power consumption is multiplied by the solutions times found for MF-CIM in~\cref{sec:results}.

The power consumption for the DL-CIM can be written as:
\begin{equation}
P_\text{max} = P_\text{opt} + 2P_\text{mod}(N-1) + P_\text{sq} + P_\text{OPA}(N-1)
\label{eq:power_DL-CCVM}
\end{equation}
Here, the first term is the power required to generate the optical pulses, the second term is the power required to run the two amplitude and phase modulators placed on each of the $(N-1)$ delay paths in the delay network, $P_\text{sq}$ is the power required to generate the injected squeezed state, $P_\text{elec}$ is the power consumed by the electronics to prepare the device for each round trip, and the last term is the power consumed by an OPA placed in the delay-line network (not shown in~\cref{fig:ccvm-schematic}) to amplify the optical pulses before sending them through the delay lines~\cite{maruo2016truncated}. \cref{tab:DL-CCVM_power} shows the parameters used for this power consumption estimation.

\begin{table}[h]
\begin{tabular}{|c|c|c|c|c|}
\hline
Term & $P_\text{opt}$ & $P_\text{mod}$ & $P_\text{sq}$ & $P_\text{OPA}$ \\
\hhline{|=|=|=|=|=|}
Value & $1.2$ mW & $10$ mW & $180$ mW & $222.2$ mW \\
\hline
\end{tabular}
\caption{The values used for the sources of power consumption in DL-CIM according to~\cref{eq:power_DL-CCVM}.}
\label{tab:DL-CCVM_power}
\end{table}

For the type of PPLN used in CIMs, the threshold power is around $270~\mu$W~\cite{mcmahon2016fully}. To obtain this power from the output of a second-harmonic generation (SHG) process with around $30\%$ efficiency, an input power of $900~\mu$W is needed. We have added another $300~\mu$W of power to this value for $P_\text{opt}$ to account for extra laser source power requirements such as homodyne detection or phase or amplitude modulators. $P_\text{mod}$ is estimated based on amplitude and phase modulators that can have a $V_\pi$ of $1$ V for a $50~\Omega$ input impedance. This is the voltage required to generate a full-range intensity or phase modulation. Modulators based on thin-film lithium niobate are capable of generating such low power modulations~\cite{xu2022dual}. $P_\text{sq}$ is estimated to be the power required for generating a maximum anti-squeezing of $24.7$ dB~\cite{shi2018detection}. An OPA is placed at the input of the delay-line network to amplify the optical pulses so that there is enough optical power going into each optical delay line to pass through the modulators. Assuming $1$ mW power required at each optical line, the output of the OPA needs to provide $(N-1)\times 1$ mW of power. Assuming a conversion efficiency of $0.015$~\cite{OPA}, and taking into account the SHG process required before the OPA, the value $P_\text{OPA} = 222.2$ mW is estimated.

\subsection{Further benchmarking analysis}
\label{sec:results}

\begin{figure*}[t]
\subfloat[\label{fig:tts-parallel}]{\includegraphics[scale=0.255]{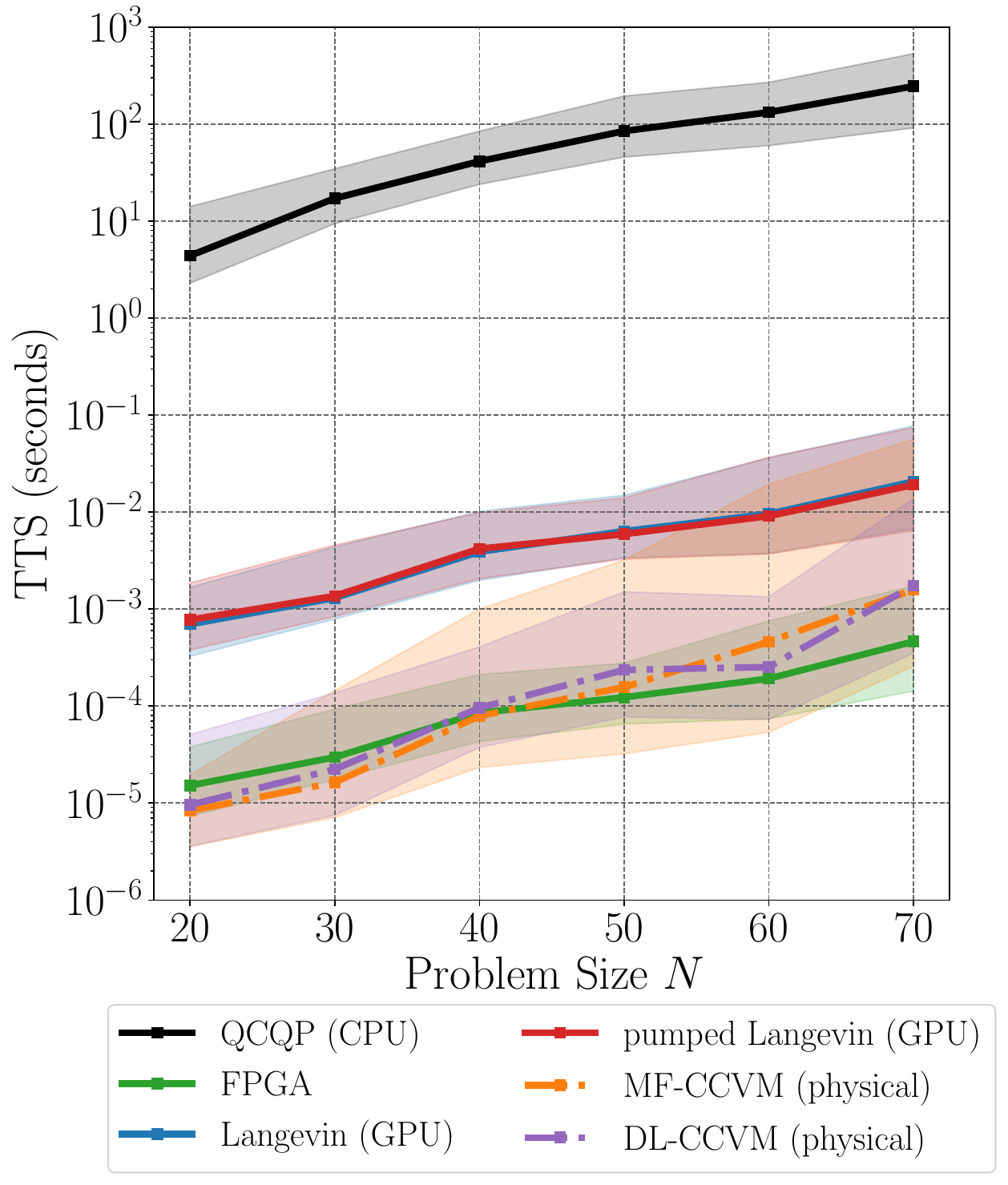}}\hspace{2mm}
\subfloat[\label{fig:ets-parallel}]{\includegraphics[scale=0.255]{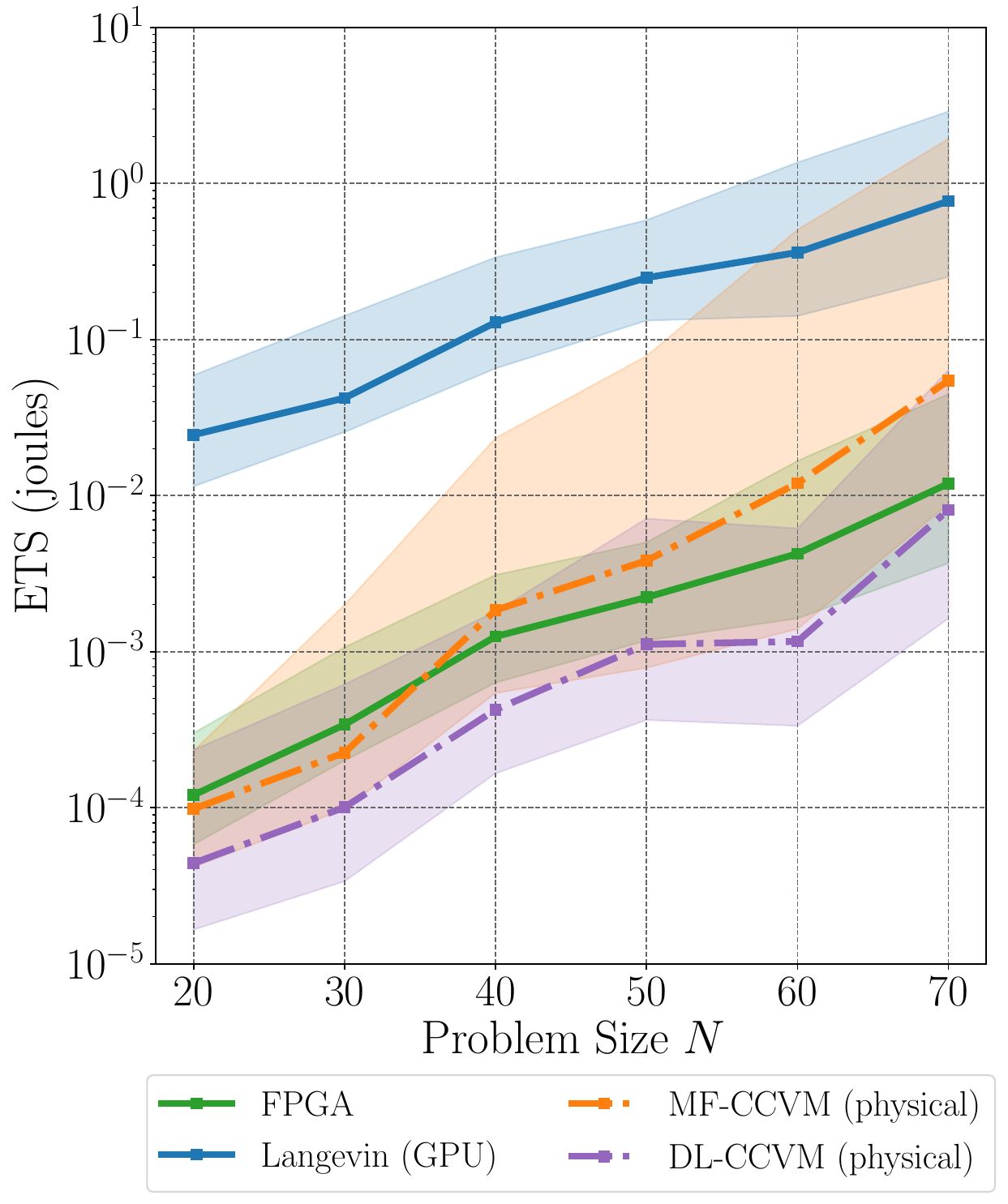}}\\
\caption{(a) Wall-clock TTS for optics-based CV impelemntations of CIM (dashed--dotted curves) in comparison with the wall-clock TTS for various classical solvers implemented on conventional digital devices (solid curves). The latter were computed based on the implementation time on the CPU, GPU, and FPGA, while the former were calculated using the parameters of the corresponding physical device. The FPGAs in the full FPGA (green curve) and MF-CIM (orange curve) schemes are designed using a \textit{parallel} implementation to solve multiple instances during the same run of the device. (b) Estimated ETS for optics-based CV implementations of CIM (dashed--dotted curves) in comparison with the ETS measured on various digital devices (solid curves). The fully optical DL-CIM implementation has the lowest energy consumption due to the absence of digital-electronic components. The plot also shows that the smaller problem instances presented here can benefit from parallelization on FPGA in the FPGA and MF-CIM schemes.
\label{fig:tts&ets-parallel}}
\end{figure*}

In this section, DL-CIM and MF-CIM are treated as solvers that can be implemented on both standard digital and analog optical devices. \cref{fig:tts&ets-parallel} shows further benchmarking comparison of the estimated physical TTS for the optics-based implementations of DL-CIM and MF-CIM, the TTS for all four solvers introduced in~\cref{sec:langevin_and_con} implemented on conventional digital devices (i.e., those based on electronics, not optics), and the TTS for the quadratically constrained quadratic programming (QCQP) solver~\cite{park2017general}, a general heuristic solver for quadratic programming problems. Our CV implementation of MF-CIM and fully FPGA-based implementations incorporate parallelization in the design of the FPGA in this case. In the MF-CIM, due to the lower speed of the FPGA compared to the optics, the large latency of the FPGA can be compensated for by performing certain computations in parallel and solving multiple problems simultaneously. Calling this a \textit{parallel} implementation, this scheme solves multiple instances of a problem during the same run of the device. Although this can lead to a larger FPGA design with higher power consumption, it can significantly improve the TTS and ETS metrics (see \cref{sec:FPGA_parallelization}). Note that the \textit{parallel} implementation in the FPGA and MF-CIM architectures as well as the GPU results are not scalable for arbitrarily large problem sizes as the computation resources scale with problem sizes and it becomes physically infeasible to implement such hardware for much larger problem sizes. Therefore, the \textit{parallel} implementation results in \cref{fig:tts&ets-parallel} should only be viewed for the problem sizes studied and not for analyzing the scaling of different types hardware. \Cref{fig:ets-parallel} shows the ETS found for the OLD solvers implemented on three digital electronic hardware devices, an FPGA, a CPU, and a GPU, as well as the estimated ETS values for the DL-CIM and MF-CIM architectures, incorporating a \textit{parallel} scheme for the FPGA and MF-CIM implementations.

As shown in~\cref{fig:tts&ets-parallel}, incorporating a \textit{parallel} implementation in FPGA and MF-CIM improves the TTS and ETS to be comparable to the DL-CIM scheme. Although a fully optical device is generally expected to outperform a hybrid optical-digital or a fully digital device, there are a few factors contributing to this comparable performance. The DL-CIM scheme in~\cref{eq:sde_DL-CIM} provides lower quality solutions compared to the OLD~\cref{eq:sde_langevin} or the MF-CIM scheme \cref{eq:sde_MF-CIM} due to the extra terms present the SDEs of the DL-CIM and the replacement of the clamp function with the implementation of the box constraint using the saturation functionality of the nonlinear crystal in the DL-CIM. Furthermore, to improve the quality of the results of all solvers, a simple post-processing step on a CPU is implemented (see~\cref{sec:tts_appndx}). Although this step does not have a significant effect on the FPGA and MF-CIM implementations, it can degrade the TTS of the DL-CIM by one order of magnitude. Additionally, the DL-CIM only incorporates a single hardware designed for the given problem size, whereas the \textit{parallel} implementation of the FPGA and MF-CIM features a specific design for each problem size. These observations indicate that the true merits of a fully optical device would become more evident for larger problem sizes and for computational tasks that are closer in nature to the SDEs of a fully optical device. In addition, having comparable results for the parallel implementations of the OLD on FPGA and the MF-CIM shows that the advantage of parallel implementation originates in the parallelization in the digital processor and not the use of optics in MF-CIM.

Together with the results in \cref{sec:the_ugly}, these results encourage research into further parallelization for performing MVM operations in a fully analog optical device, as opposed to time-multiplexed MVM in CIM, to further improve TTS and as a consequence ETS metrics of a fully optical device. In addition, having reliable, energy-efficient, and fully analog nonlinear function units can greatly improve the performance of an optical solver when implementing nonlinear functions such as the box constraint implemented in this paper. Various architectures have been proposed for implementing nonlinear functions in optics~\cite{xu2022reconfigurable,mourgias2019all-optical,li2023all-optical}.

\section{Parallelization in the FPGA\label{sec:FPGA_parallelization}}

In this paper, we have investigated two possible implementations for the FPGA in the OLD and the FPGA in the MF-CIM. The \textit{serial} implementation uses the smallest computation unit for performing a single multiplication and addition  at a time. Although in practice multiple operations can be performed simultaneously on FPGA, this approach provides realistic curves for the TTS and ETS that are applicable to large problem sizes.

Here, we have also investigated a \textit{parallel} implementation of the FPGA which can lead to improvements in the TTS and ETS of the OLD implemented on FPGA and the MF-CIM. In the MF-CIM, the latency due to the FPGA and the ADC and DAC is the main bottleneck in the computation of the drift term during each iteration, due to the clock speed of the FPGA (here being $300$ MHz) being much lower than the pulse rate of the laser (here being $5$ GHz). As a consequence, the optical components idle while waiting for the FPGA to generate the drift amplitudes and, additionally, only parts of the FPGA will be actively used to perform the computation at a given time, leading to the FPGA idle at some points and consequently having larger TTS and ETS in MF-CIM. To take advantage of the full FPGA architecture at all times, similar to the GPU implementation, we can solve multiple instances of a given problem simultaneously. In this case, the pulse train would be multiplied to represent multiple copies of the variables in order to solve the same problem instance many times simultaneously. The number of times a given instance would be solved at a given run of the machine is denoted by the batch size.

\begin{figure}[!t]
\centering
\subfloat[\label{fig:fpga_MVM}]{\includegraphics[width=0.9\linewidth]{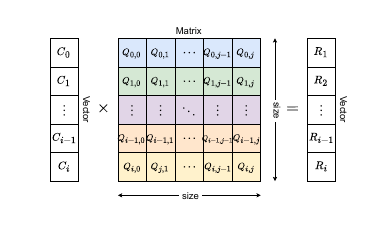}}\\
\subfloat[\label{fig:fpga_serial}]{\includegraphics[width=1.0\linewidth]{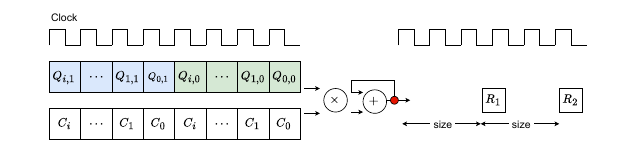}}\\
\subfloat[\label{fig:fpga_parallel}]{\includegraphics[width=1.0\linewidth]{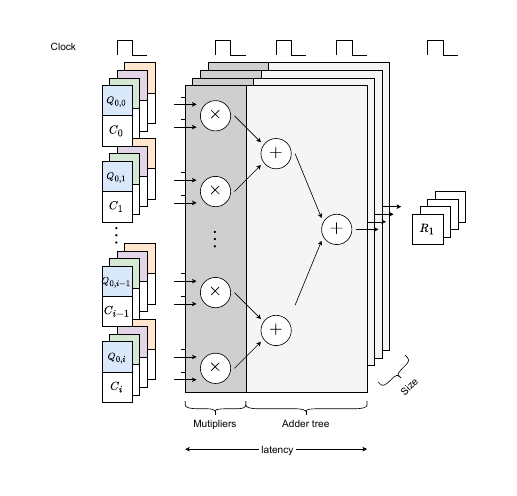}}\\
\caption{The design of the FPGA used for the TTS and ETS estimations in this paper: (a) The variables used in the process of MVM, (b) \textit{serial} implementation of MVM using a single ``compute unit", and (c) \textit{parallel} implementation of the FPGA using multiple ``compute units" simultaneously.\label{fig:fpga_architecture}}
\end{figure}

\begin{figure}[!t]
  \subfloat[\label{}]{\includegraphics[scale=0.215]{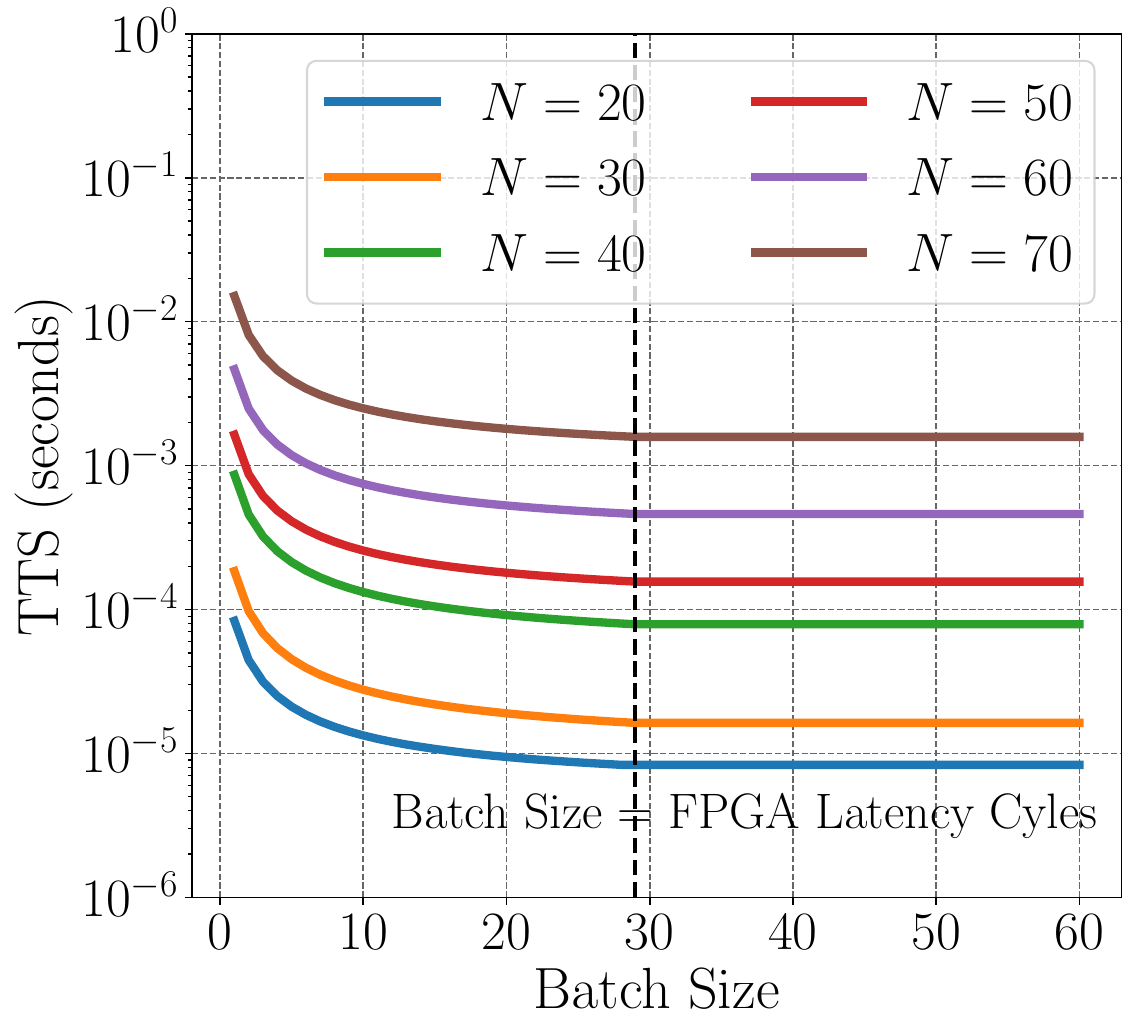}}\hspace{2.0mm}
  \subfloat[\label{}]{\includegraphics[scale=0.215]{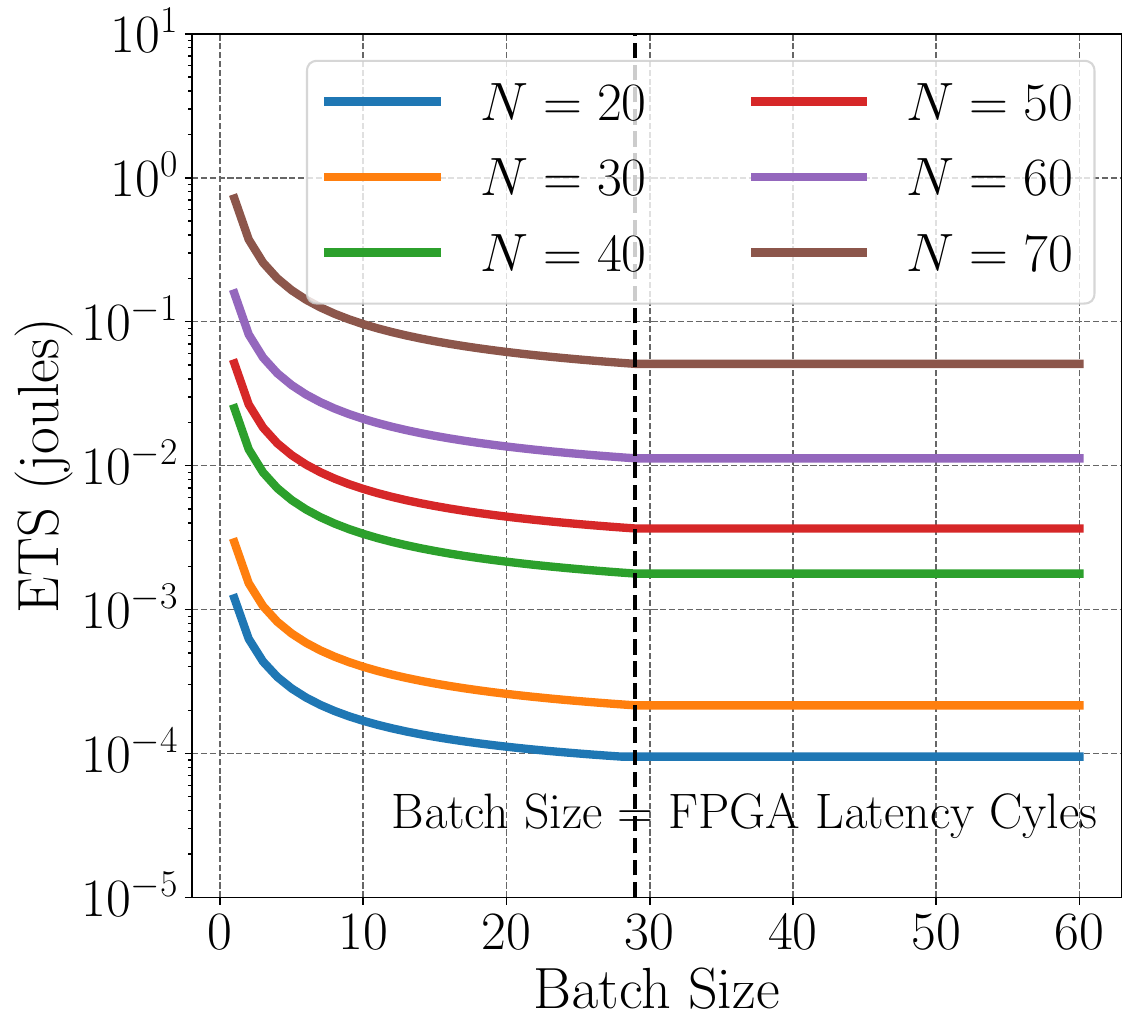}}
  \caption{The results for the MF-CIM versus the batch size for (a) physical TTS and (b) ETS, plotted for problem sizes $20$ to $70$. TTS and ETS improve with increasing the batch size up to the batch size of $29$; this is the value where the number of times an instance is solved simultaneously is equal to the latency cycles of the FPGA, maximizing the use of its architecture.\label{fig:tts&ets_vs_batchsize}}
\end{figure}

\cref{fig:fpga_architecture} shows the FPGA MVM design for the \textit{serial} and \textit{parallel} implementations. \cref{fig:fpga_MVM} shows the variables and the parameters used in an MVM operation. \cref{fig:fpga_serial} shows the architecture of the FPGA for the \textit{serial} implementation. This structure consist of having a single “compute unit” which sequentially multiplies and accumulates the result of each column of the given matrix and provides the result after the operation is completed.  This implementation uses fewer resources but takes more time to complete.
\cref{fig:fpga_parallel} shows the architecture of the FPGA for the \textit{parallel} implementation. Leveraging both row and column parallelism, the computation is highly optimized by simultaneously processing multiple elements. Row parallelism involves distributing each row of the matrix across different processing units, allowing for simultaneous multiplication of all the elements in a row by their corresponding vector elements. Column parallelism, on the other hand, distributes the vector across processors so that each element of the vector is multiplied with every element of a matrix column at the same time. This dual approach maximizes the use of available hardware resources, significantly speeding up the computation by performing many operations in parallel. The combination of row and column parallelism ensures  that the entire MVM is completed in a fraction of the time it would take with the \textit{serial} implementation.

\cref{fig:tts&ets_vs_batchsize} shows the TTS and ETS for the MF-CIM as a function of the batch size for different instance sizes when implementing the \textit{parallel} scheme. Evidently, both metrics improve with increasing the batch size, up to the batch size of $29$ where the values reach a plateau. This is the point where the batch size is equal to number of cycles of the FPGA's latency. At this point, all of the FPGA is being used to perform calculations simultaneously and no parts of it are idling. This approach is used for the parallel implementation of the OLD on the FPGA as well. The ETS and TTS values shown in \cref{fig:tts&ets-parallel} are evaluated using this method. Note that the values presented for the \textit{parallel} implementations are only applicable to the problem sizes implemented here and should not be used to estimate the TTS and ETS for larger problem sizes as the computational resources become limited when the problem size increases.

\section{Time evolution of the CIM dynamics
\label{sec:time-evolution}}

\begin{figure}[!h]
\subfloat[\label{fig:time-evol-instance1}]{\includegraphics[scale=0.215]{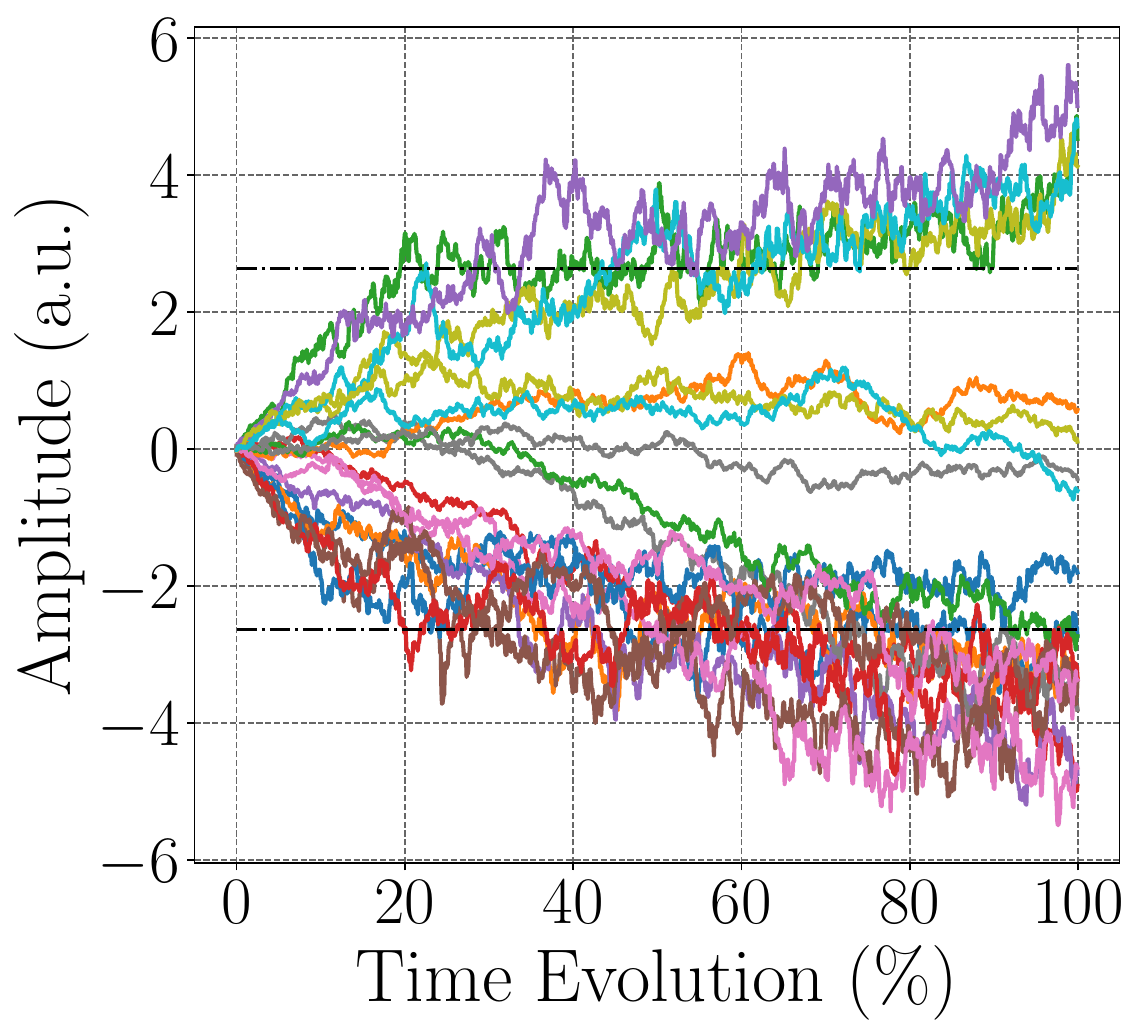}}\hspace{2mm}
\subfloat[\label{fig:time-evol-instance2}]{\includegraphics[scale=0.215]{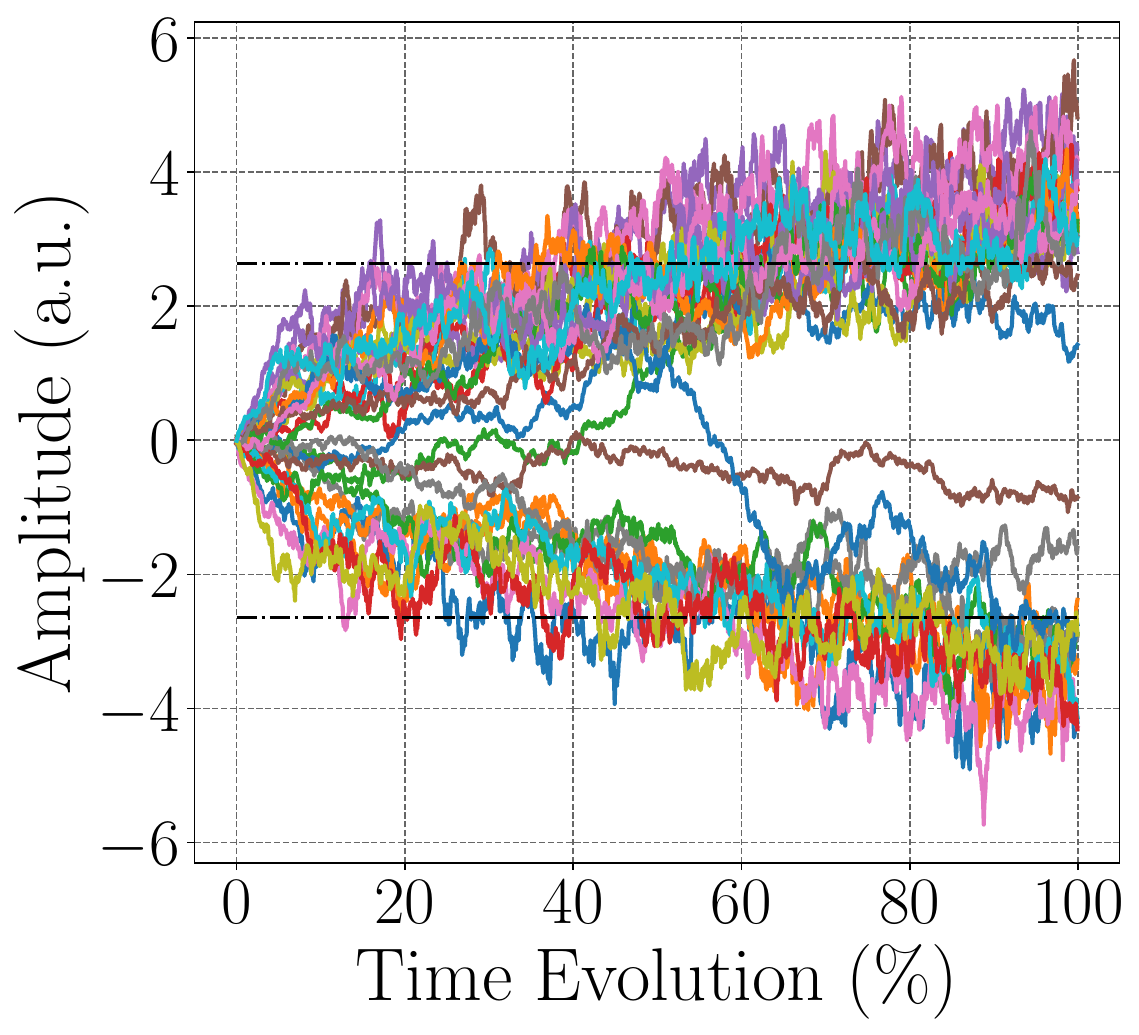}}\\
\subfloat[\label{fig:time-evol-instance3}]{\includegraphics[scale=0.215]{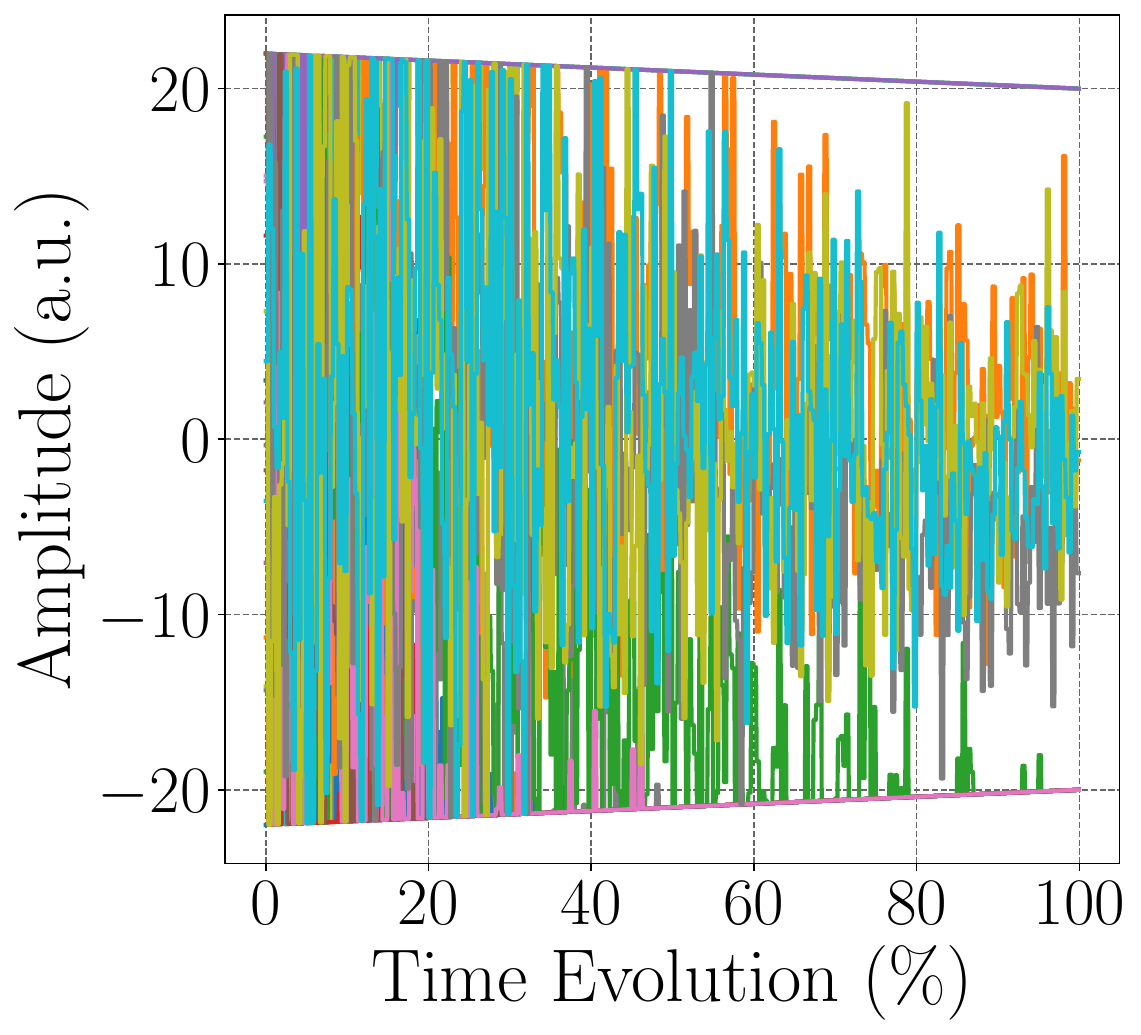}}\hspace{2mm}
\subfloat[\label{fig:time-evol-instance4}]{\includegraphics[scale=0.215]{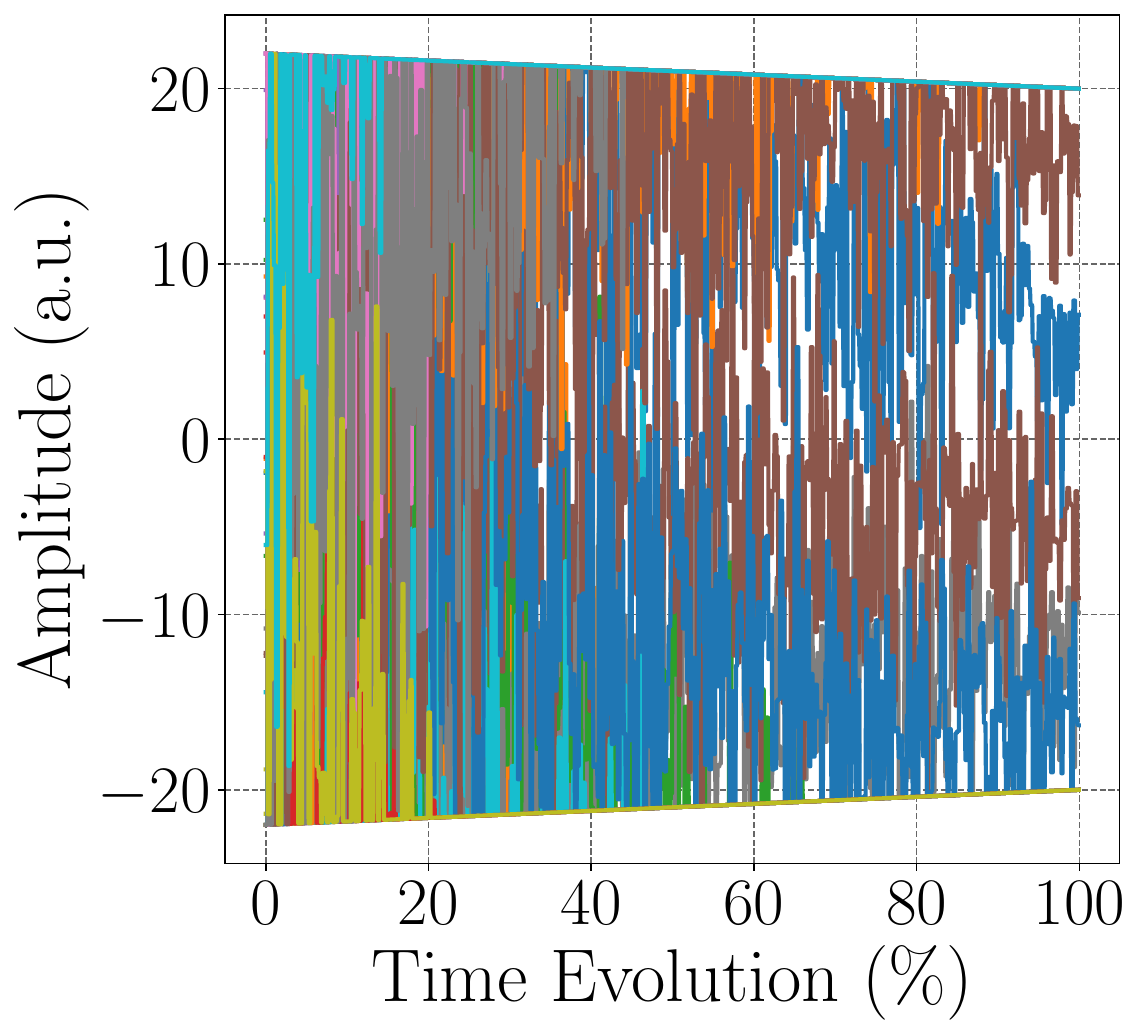}}
\caption{Time evolution of the DOPO pulses' amplitudes for four randomly generated BoxQP problem instances. A problem of size (a) $20$ and (b) $30$ solved using CV readout of the DL-CIM, and the same problems of size (c) $20$ and (d) $30$ solved using CV readout of the MF-CIM. The heavy black dashed lines in panels (a) and (b) indicate the values $-s$ and $s$, at which the amplitudes are clamped prior to computing the objective function. In panels (c) and (d) for the MF-CIM, the value at which the pulses are clamped at each round trip is linearly scheduled to reach $s=20$ at the end of the evolution process. For both solvers, the objective function is calculated after the substitution $x_i = \frac{1}{2}(c_i/s + 1)$ in the DL-CIM or $x_i = \frac{1}{2}(\tilde{\mu}_i/s + 1)$ in the MF-CIM, to satisfy the box constraint in~\cref{eq:boxQP}, with $l_i = 0$ and $u_i = 1$ for all $i\in\{1,\ldots,N\}$. Note that, for each of the problem instances, the majority of the amplitudes have an  optimal value at the boundary of the box constraint, while only some of the amplitudes converge to fractional values.}
\label{fig:time-evolution_DL-CIM}
\end{figure}

\cref{fig:time-evolution_DL-CIM} shows the time evolution of the DOPO pulses' amplitudes for the DL-CIM and MF-CIM dynamics~(see \cref{eq:sde_DL-CIM,eq:sde_MF-CIM}). The heavy dashed lines in \cref{fig:time-evol-instance1,fig:time-evol-instance2} indicate the values $-s$ and $s$ at which the amplitudes are clamped at the end of the process in the DL-CIM. Since the clamping is applied at the end of the evolution, it can be considered a post-processing step that is performed on a classical computer. In the \mbox{MF-CIM} time evolution plots in~\cref{fig:time-evol-instance3,fig:time-evol-instance4}, the values at which each pulse is clamped at each round trip is linearly scheduled to reach a value of $20$ at the end of the process. The system parameters, including the pump value, are optimized for the highest success probability through parameter tuning. Their ranges are shown in~\cref{tab:solvers_parameters}.
At the end of the evolution process, the pulse amplitudes are plugged into the equation $x_i = \frac{1}{2}(y_i/s + 1)$, where $y_i = c_i$ for the DL-CIM and $y_i=\tilde{\mu}_i$ for the MF-CIM, to obtain the problem variables in order to implement the box constraint $0\le x_i \le 1$. It is evident that, for the solutions of the problems, the majority of the variables are found to be at the edges of the box constraint at $x_i = 0$ or $x_i = 1$, as explained in~\cref{sec:results,sec:app-instance_generation}.

\section{Choosing the number of iterations}
\label{sec:tts_vs_iter}

\begin{figure}[thb]
  \subfloat[\label{}]{\includegraphics[scale=0.215]{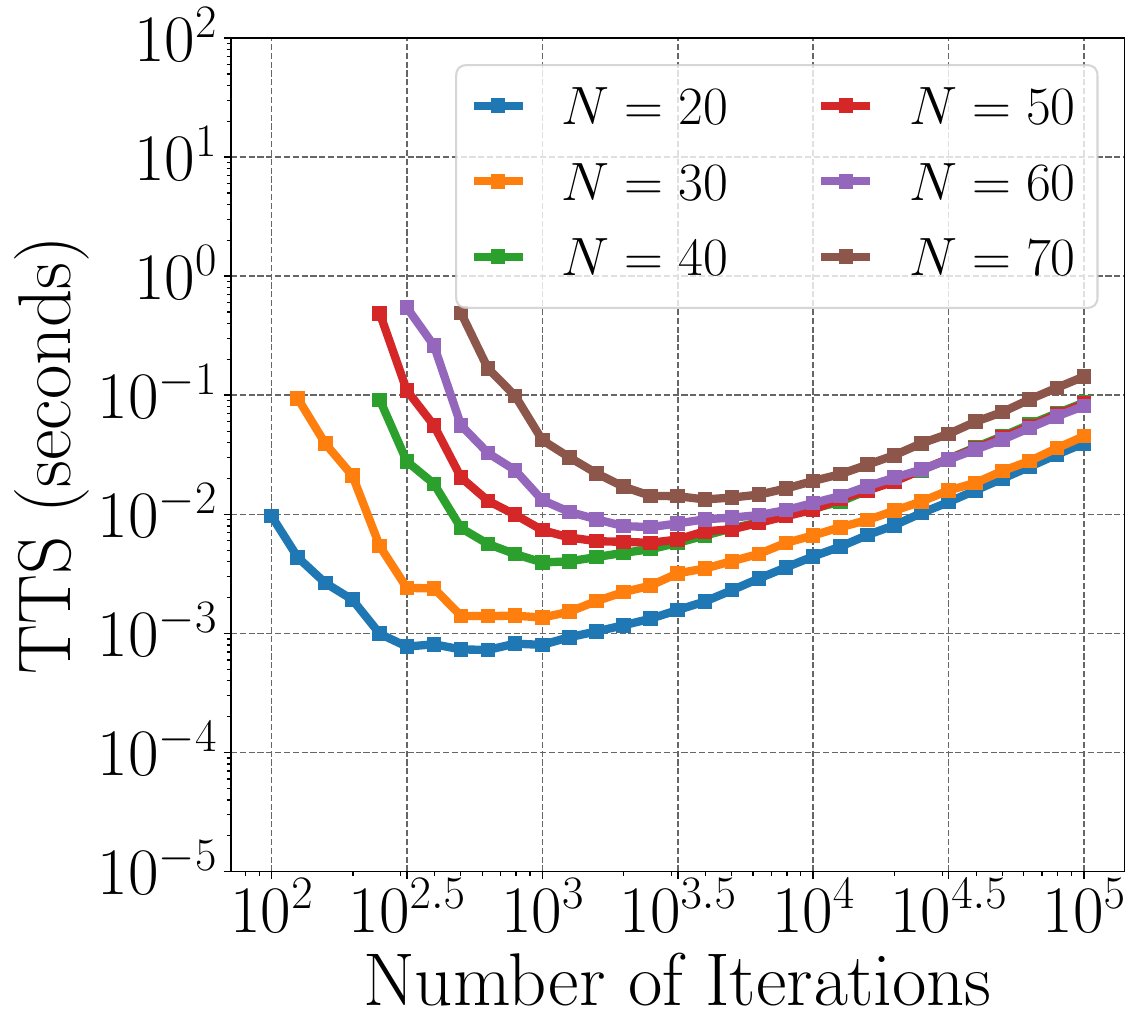}}\hspace{2.0mm}
  \subfloat[\label{}]{\includegraphics[scale=0.215]{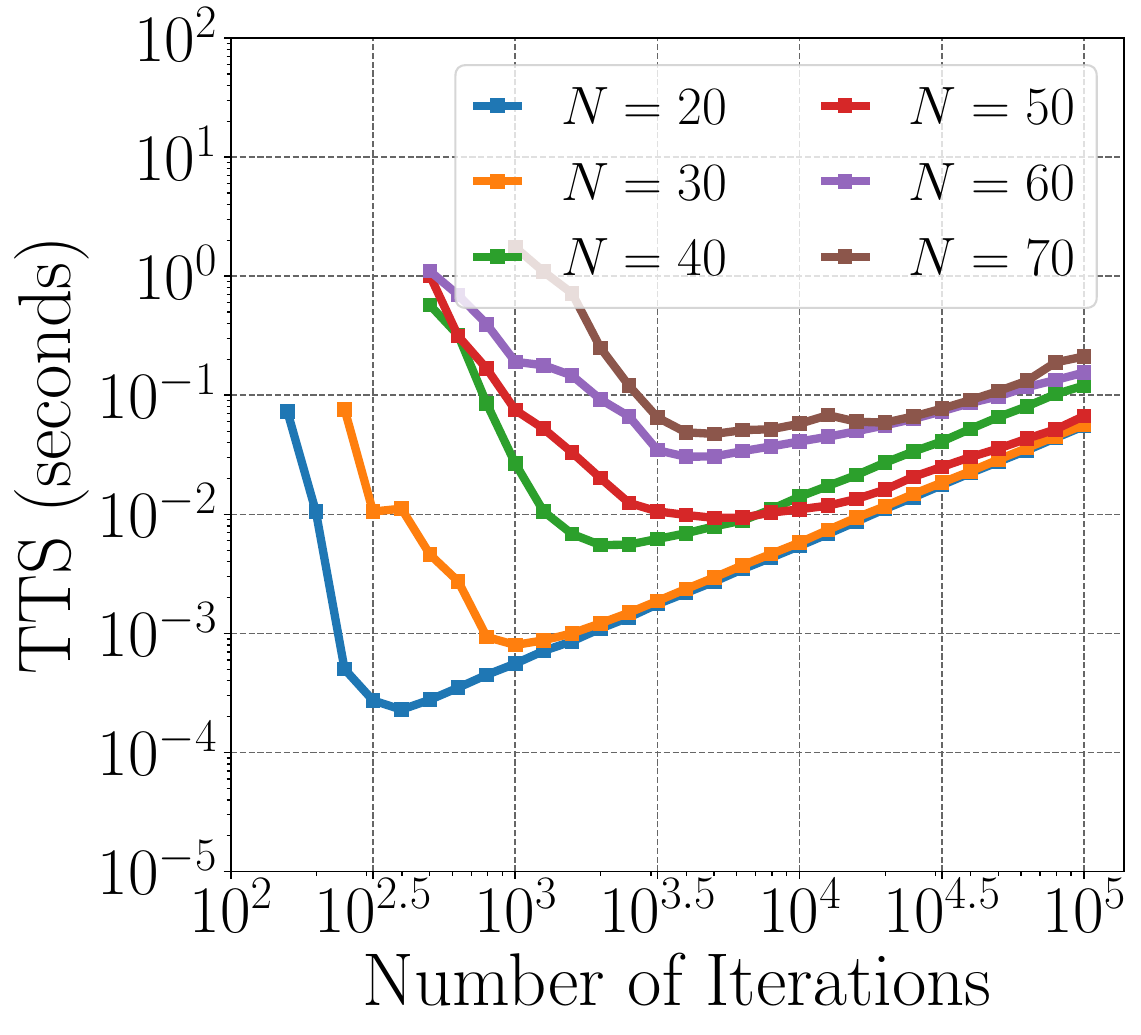}}\\
  \subfloat[\label{}]{\includegraphics[scale=0.215]{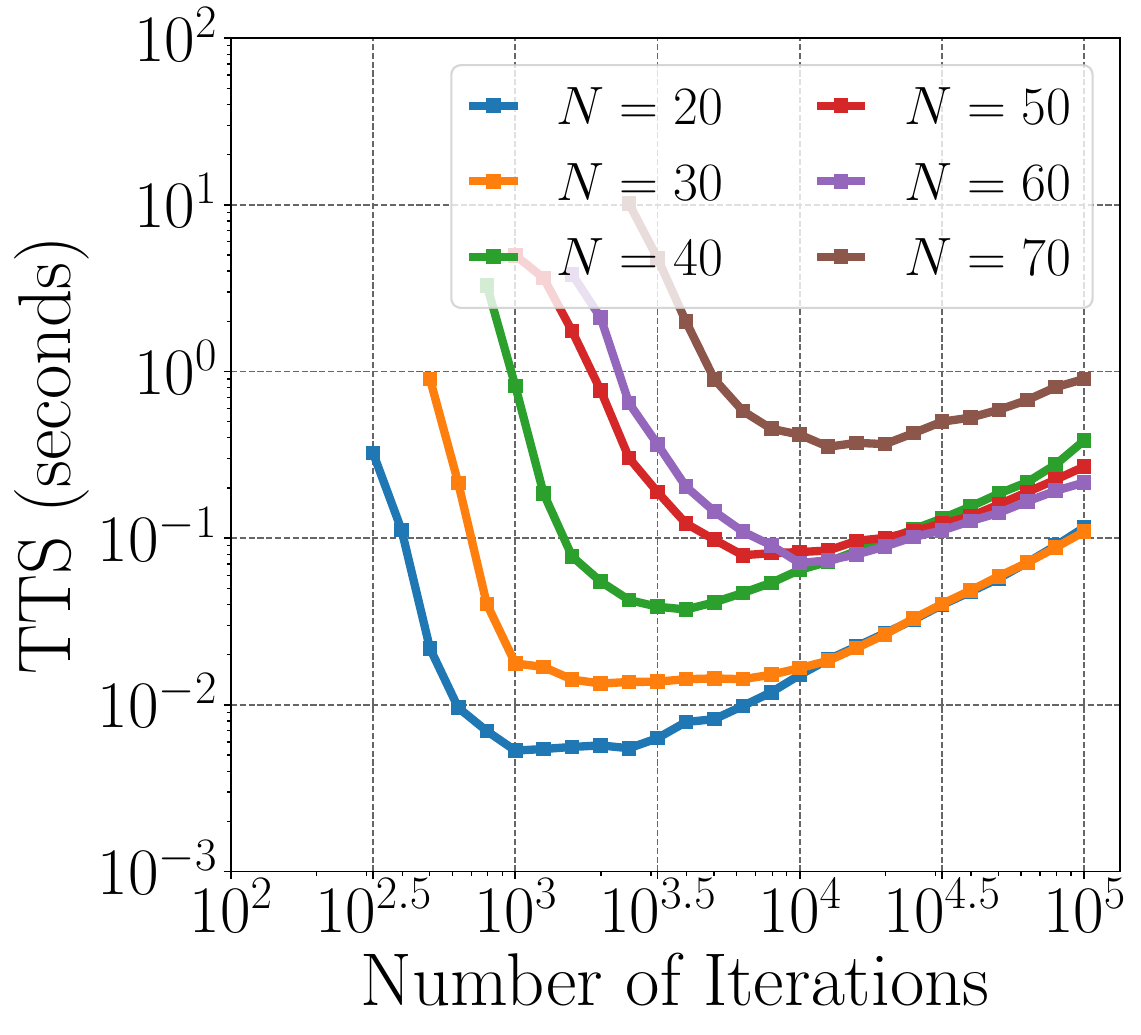}} \hspace{2.0mm}
  \subfloat[\label{}]{\includegraphics[scale=0.215]{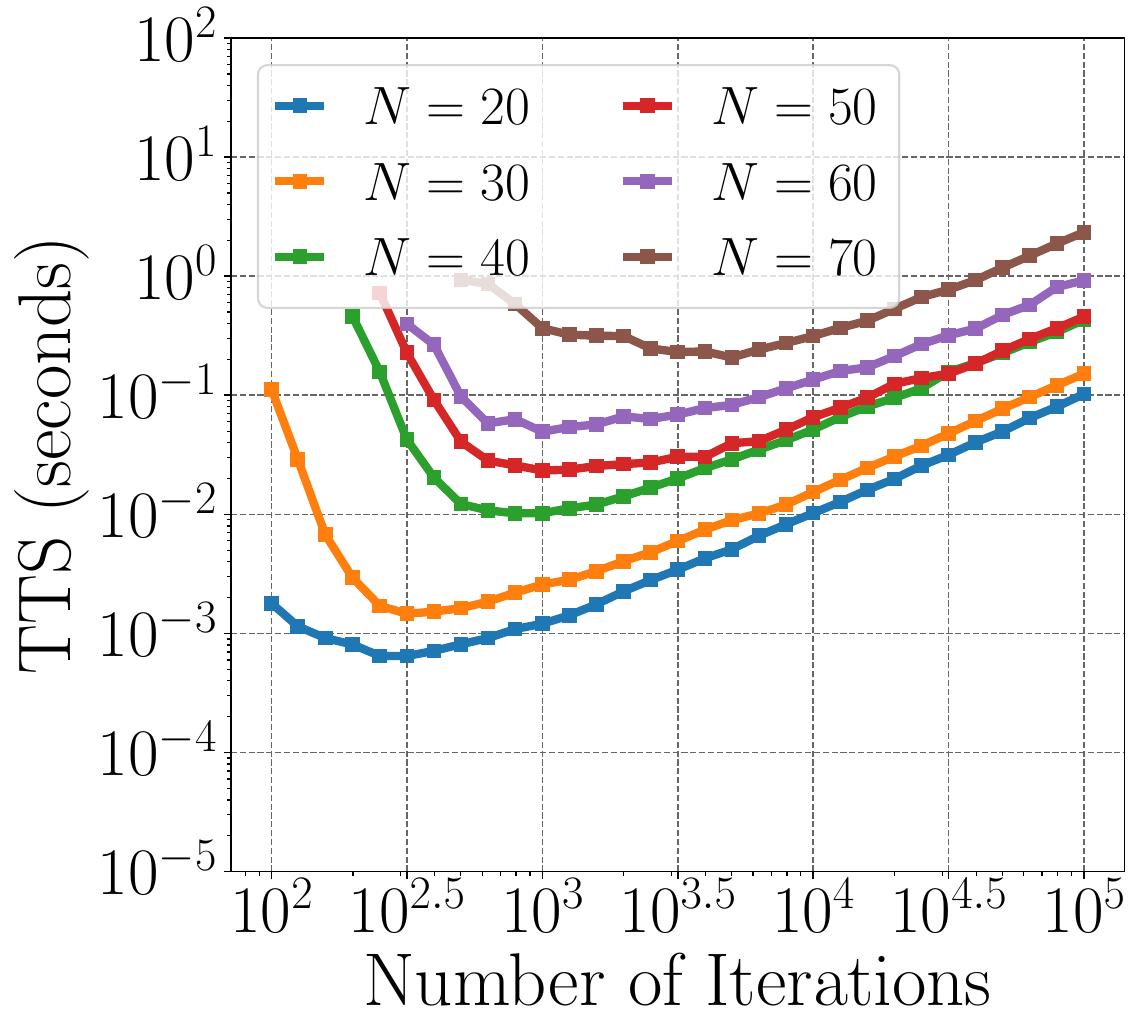}}
  \caption{TTS versus number of iterations for different problem sizes for (a) OLD, (b) PLD, (c) DL-CIM, and (d) MF-CIM. \label{fig:tts-vs-iter}}
\end{figure}

\cref{fig:tts-vs-iter} shows the dependency of the TTS on the number of iterations for each solver. For each problem size, there is an $n_\text{iter}$ that gives the lowest TTS. This is because for $n_\text{iter}$ larger than a certain value, the success probability does not significantly improve but TTS gets larger due to the increase in the number of round-trips. We have used the curves in \cref{fig:tts-vs-iter} to find the optimum $n_\text{iter}$ for each problem size for each solver and plot the TTS and ETS plots in the main body of the paper.

\clearpage
\bibliography{refs}

\end{document}